%% file: main.tex
\documentclass[11pt]{article}

\usepackage[a4paper]{geometry}
\usepackage{authblk}
\usepackage{amssymb}
\usepackage{array}
\usepackage{mathrsfs}
\usepackage{amscd}
\usepackage{amsmath}
\usepackage{amsfonts}
\usepackage{amsthm}
\usepackage{psfrag}
\usepackage{textcomp}
\usepackage{bm}
\usepackage{dsfont}
\usepackage{algorithm}
\usepackage{algpseudocode}

 \usepackage{lineno}

\usepackage{amssymb}

\usepackage[figuresright]{rotating}
\usepackage{moreverb}
\usepackage{amssymb}
\usepackage{array}
\usepackage{mathrsfs}
\usepackage{amscd}
\usepackage{amsmath}
\usepackage{amsfonts}
\usepackage{amsthm}
\usepackage{psfrag}
\usepackage{textcomp}
\usepackage{bm}
\usepackage{color}
\usepackage{float}
\usepackage[latin1]{inputenc} 	
\usepackage{afterpage}
\usepackage{tabularx}
\usepackage{booktabs}
\usepackage{multirow}
\usepackage[normalem]{ulem} 
\usepackage{cancel}
\usepackage{graphicx}

\usepackage[square, numbers, comma, sort&compress]{natbib}

\usepackage{setspace}
\usepackage[colorlinks]{hyperref}
\usepackage{xcolor}
\hypersetup{
    colorlinks,
    linkcolor={lightblue},
    citecolor={lightblue},
    urlcolor={lightblue}
}

\definecolor{lightblue}{RGB}{6,69,173}

\def\XM {{  x_{\alpha} -  \Delta x_{\alpha} /2 }}

\def\XP {{ x_{\alpha} +  \Delta x_{\alpha} /2  }}

\def\XMfrac {{  x_{\alpha} -  \frac{\Delta x_{\alpha}}{2} }}

\def\XPfrac {{ x_{\alpha} +   \frac{\Delta x_{\alpha}}{2}   }}

\def\Dxa {{ \Delta x_{\alpha} }}

\providecommand{\keywords}[1]
{
  \small	
  \noindent \textbf{Keywords:} #1
}
\title{Analytical approximations for multiple scattering in one-dimensional waveguides with small inclusions}

\author[1,2]{Mario L\'azaro}
\author[2]{Richard Wiltshaw}
\author[2]{Richard V. Craster}
\author[1]{Luis M. Garc\'ia-Raffi}

\affil[1]{Instituto Universitario de Matem\'atica  Pura y Aplicada, 
Universitat Polit\`ecnica de Val\`encia, 46022 (Spain)}
\affil[2]{Department of Mathematics, Imperial College London, London, SW7 2AZ, UK}

\date{}

\begin{document}

\setstretch{1.2}
\maketitle
\setstretch{1.0}
\begin{abstract}

We propose a new model to approximate the wave response of waveguides containing an arbitrary number of small inclusions. The theory is developed to consider any one-dimensional waveguide (longitudinal, flexural, shear, torsional waves or a combination of them by mechanical coupling), containing small inclusions with different material and/or sectional properties. The exact problem is modelled through the formalism of generalised functions, with the Heaviside function accounting for the discontinuous jump in different sectional properties of the inclusions. For asymptotically small inclusions, the exact solution is shown to be equivalent to the Green's function. We hypothesize that these expressions are also valid when the size of the inclusions are small in comparison to the wavelength, allowing us to approximate small inhomogeneities as regular perturbations to the empty-waveguide (the homogeneous waveguide in the absence of scatterers) as point source terms. By approximating solutions through the Green's function, the multiple scattering problem is considerably simplified, allowing us to develop a general methodology in which the solution is expressed for any model for any elastic waveguide. The advantage of our approach is that, by expressing the constitutive equations in first order form as a matrix, the solutions can be expressed in matrix form; therefore, it is trivial to consider models with more degrees of freedom and to arrive at solutions to multiple scattering problems independent of the elastic model used. The theory is validated with two numerical examples, where we perform an error analysis to demonstrate the validity of the approximate solutions, and we propose a parameter quantifying the expected errors in the approximation dependent upon the parameters of the waveguide.

\end{abstract}

\hspace{10pt}






\section{Introduction}

Phononic metamaterials have become particularly important in recent decades, motivated by the possibility of controlling the propagation of low frequency waves through structured media containing arrays  of single-frequency \cite{WangG-2005} or multiple-frequency \cite{Miranda-2019,WangZ-2013,Xiao-2012c,Xiao-2013,wiltshaw2023analytical,Cai-2022} resonators. The metamaterial design paradigm is a popular approach across all areas of wave physics to create subwavelength devices by utilising these resonances, for instance in photonics \cite{zheludev2012metamaterials,monticone2017metamaterial,ali2022metamaterials}, or phononics \cite{cummer2016controlling,brule2020emergence,muhammad2022photonic,krushynska2023emerging} where examples include energy harvesting \cite{chen2014metamaterials,chaplain2020topological,de2020graded,de2021graded} or seismic protection devices \cite{colombi2016seismic,colombi2016forests,lott2020evidence,mu2020review}.  Additionally, new configurations based on attached Rayleigh beams consider the coupling between longitudinal and flexural components of motion to control the wave response \cite{Movchan-2022a,Movchan-2022b,Movchan-2023}. In the case of flexural waves in plates, the introduction of point oscillators requires the use of multiple scattering methods together with the plane wave expansion method \cite{wiltshaw2023analytical,Torrent-2012,Wiltshaw-2020,Movchan-2017a,Movchan-2018,Ruzzene-2017}.\\

The analysis of heterogeneous materials, especially concerning their inherent resonances and properties post perturbation, is a topic of great practical interest in the fields of  experimental modal analysis or structural health monitoring. Analytical approximations in combination with finite element based methods have been proposed for vibrating structures with varying properties along their length \cite{LeeJ-2016,Adamek-2015,Aya-2012,ZhangK-2017,wang2015locally,aguzzi2022octet}, or for the evaluation of transmission and reflection due to the introduction of multiple oscillators \cite{WuJ-1998,Brennan-1999,Mace-2007,TanC-1998}. Additionally, heterogeneity models generated by internal cracks using rotational springs \cite{Krawczuk-1994a,Krawczuk-1996,Krawczuk-2001a,Krawczuk-2003b} or specially designed finite elements \cite{Yeung-2019} have been proposed. In the mechanics of nanomaterials, models based on material contrast have been proposed to evaluate the response of structures with multiple cracks distributed along their length, and their influence on the modal parameters - both for flexural waves (nanobeams) \cite{Loghmani-2018a}, and for longitudinal waves (nanorods) \cite{Loghmani-2018b}. \\

Currently, a multitude of analytical theories exist to accurately compute the dispersion relation of elastic homogeneous waveguides. These theories ~\cite{Doyle-1997,Graff-1999,Royer-2022} are valid for wavelengths up to the order of magnitude of the cross section, thereby limiting their applicability to a few particular geometries. The most popular analytical method to calculate  the response of one-dimensional waveguides with piecewise property changes is the Transfer Matrix Method (TMM) \cite{Rui-2019}. This approach obtains the analytical solution in structures formed by sections and materials of different nature \cite{Chouvion-2010,Chouvion-2011,Rui-2019,WangZ-2013,Lazaro-2022a} at the expense of introducing some practical limitations: firstly, the properties need to be defined piece-wisely. Therefore, if sectional and/or material changes are considered in small segments of a host material (inclusions), and the number of these is high, then the analytical treatment can be tedious for obtaining the response. Secondly, it is known that the presence of evanescent waves in the model (e.g. as in Euler-Bernoulli beams) produces numerical instabilities, which are more pronounced when the propagation length is large or when there are many scatterers. \\

Moreover, for higher frequencies, theories derived from Lamb and Rayleigh waves become necessary and the analytical approach to the problem becomes increasingly complex - as more degrees of freedom are required to appropriately model a waveguide in the high frequency limits, as shown by the evolution of rod or beam theories. The complexity gained with increasing degrees of freedom makes the analytical solutions associated with the differential equation for time harmonic motion cumbersome, especially when considering the presence of localized disturbances along the medium - such as point like attached objects or differing material and/or cross-sectional heterogeneities. Therefore, it is advantageous to express the constitutive equations in first order form \cite{langley1999wave}, as the continuity conditions for ever increasing degrees of freedom (displacement, rotation, force, moments, $\ldots$) are expressed simply as a matrix. \\

 Subsequently, the aim of this paper is to develop a general analytical framework in matrix form to determine the wave response of one-dimensional elastic waveguides containing an arbitrary number of scatterers (inclusions or heterogeneities within the waveguide), and modelled assuming an arbitrary number of degrees of freedom. The expressions developed readily account for the higher degrees of freedom required by more complex models. The scatterers are assumed to be homogeneous along their length, and account for any changes in mechanical and/or sectional properties along the waveguide. We show how any such scatterer can be modelled using the Heaviside function and how, when the length of the scatterer is small, the solution closely relates to the Green's function. The results of the approximation are checked against the exact solutions derived from the TMM for two systems: a rod (classical longitudinal waves) and a Timoshenko beam (flexural waves). The method is shown to be readily applied to consider models with higher-order degrees of freedom, and is validated in the frequency range compatible with the structural model under consideration. Lastly, we propose a parameter dependent upon the material parameters of the waveguide to estimate the expected errors of the approximation used.

~
\\

\section{Elastic waveguides with a distribution of multiple scatterers}

In this section, we develop a general methodology to model one-dimensional waveguides with an arbitrary number of degrees of freedom, formed from a homogeneous material containing an arbitrary number of homogeneous inclusions with different material and/or sectional properties to the host medium. We derive general expressions for scattering from these inclusions appropriate for any one-dimensional elastic waveguide. We show that, when the inclusions are asymptotically small these solutions are equivalent to the Green's function; moreover, when the inclusions are small in comparison to the wavelength we hypothesize that this approximation is still valid, and can easily be used to simulate multiple scattering problems in any one-dimensional waveguide. \\

One-dimensional models allow us to express the displacement field of any cross-section as function of generalized variables which only depend on the longitudinal coordinate $x$. Hamilton's principle, together with the kinematic assumptions, leads to a set of partial differential equations in space-time variables $(x,t)$ written in terms of $m$ kinematic variables and $m$ generalised forces - denoted by $\bm{v}(x,t)$ and $\bm{F}(x,t)$ respectively. In general, all variables of the system can be expressed by a column vector, the so-called state--vector $\bm{u}(x,t)$ of size $2m$ as follows
\begin{equation}
		\bm{u}(x,t) = \left\{\begin{array}{cc} \bm{v}(x,t)  \\ \bm{F}(x,t) \end{array} \right\} .
		\label{eq000}
\end{equation}
Assuming that the waveguide is homogeneous and has a time harmonic response, i.e. that $\bm{u}(x,t) = \mathbf{u}(x) e^{i \omega t}$ for radian-frequency $\omega$, the space-time partial differential equations governing the constitutive relations  can be expressed as
\begin{equation}
	\frac{\textrm{d} \mathbf{u}}{\textrm{d} x} = \mathbf{A} \, \mathbf{u} + \mathbf{f}(x) \, ,
		\label{eq001}
\end{equation}
where the matrix $\mathbf{A}$ is a $(2m) \times (2m)$ matrix with frequency dependent terms containing the stiffness, mass and inertia parameters of the waveguide. The vector $\mathbf{f}(x)$ represents the distributed external loads associated with any generalized forces acting over the waveguide, and hence $\mathbf{f}(x)$ only has entries in the last $m$ terms. \\

In Table \ref{tab01} the state--vector, the external forces  and the matrix $\mathbf{A}$  are listed for two particular cases: a rod (longitudinal classical waves) and a Euler-Bernoulli beam. More examples of 1D waveguides covering other models of longitudinal waves, torsional waves, flexural waves and their coupling are shown in \ref{ap_C}. \\
\begin{table}[h]	
\hspace*{-0.25cm}
\scalebox{0.9}{
		{\input{table_01.tex}}	
  }
		\caption{The state--vector, force vector and matrix of parameters for two widely-used elastic waveguides: longitudinal waves (classical rod), flexural waves (Euler-Bernoulli beam). More examples of elastic waveguides have been listed in Table \ref{tab03} in \ref{ap_C}.}
	\label{tab01}
\end{table}

\begin{figure}[h]%
\hspace*{-0.25cm}
		\begin{tabular}{c}
			\includegraphics[width=15cm]{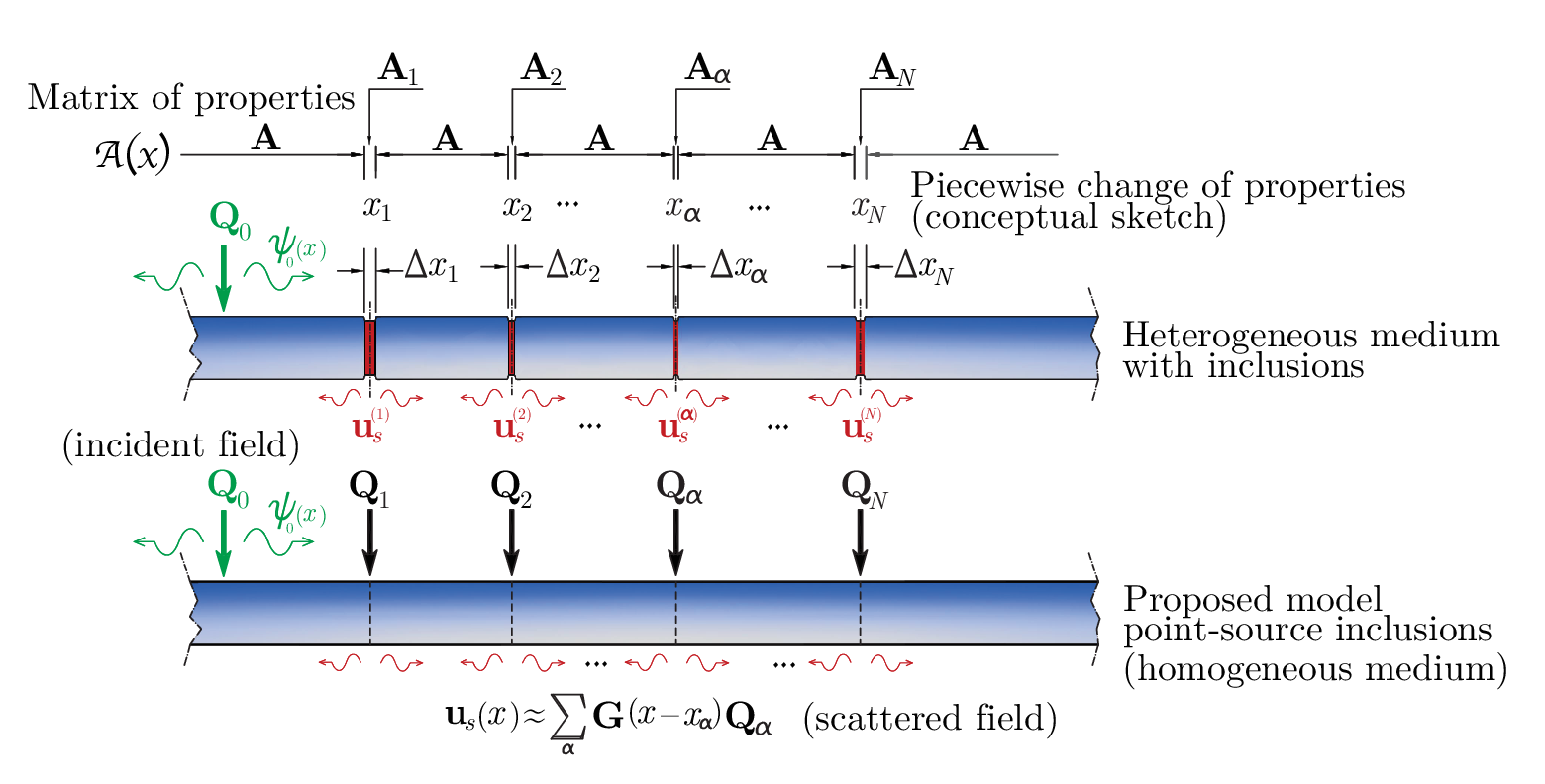} 
		\end{tabular}			
		\caption{A distribution of scatterers along a one-dimensional elastic waveguide. Here we show how the matrix $\bm{\mathcal{A}}$ in Eq.~\eqref{eq010} behaves along the waveguide (top), an example waveguide of consideration (middle) and the point-source approximation of the waveguide (bottom).}%
		\label{fig01}%

\end{figure}

We introduce a set of $N$ inclusions embedded throughout the elastic waveguide, these inclusions are assumed to be homogeneous sections with different material/sectional properties to the host medium. We shall denote parameters belonging to the inclusion $\alpha$ by subscript $\alpha$, where $\alpha$ takes values $1$ to $N$ to enumerate the inclusion considered, and we consider the general case where the inclusions can have different parameters to one another as in Fig.~\ref{fig01}. We denote the length of each inclusions $\Delta x_{\alpha}$, whose constitutive relations are completely described by the matrix $\textbf{A}_{\alpha}$ inside the interval $x_\alpha - \Delta x_\alpha/2 < x < x_\alpha + \Delta x_\alpha/2$. \\

Therefore, the differential equation governing the state--vector along the waveguide is
\begin{equation}
	\frac{\textrm{d} \mathbf{u}}{\textrm{d} x} = \bm{\mathcal{A}}(x) \, \mathbf{u} + \mathbf{f}(x) \, ,
\label{eq010}
\end{equation}
where the function $\bm{\mathcal{A}}(x) $ can be defined as the piecewise continuous function 
\begin{equation}
	\bm{\mathcal{A}}(x) =
	\begin{cases}
		\mathbf{A}_\alpha & \text{if \ }  x  \in \left[ x_\alpha - \Delta x_\alpha  / 2 , x_\alpha  + \Delta x_\alpha / 2 \right]   \\
		\mathbf{A} 				 &	 \text{if \ } x \notin \left[ x_\alpha - \Delta x_\alpha  / 2 , x_\alpha  + \Delta x_\alpha / 2 \right] 
	\end{cases} \quad \alpha = 1, 2, \ldots, N.
   \label{eq011a}
\end{equation}
This function can be expressed mathematically in one single line as
\begin{equation}
	\bm{\mathcal{A}}(x) = \mathbf{A} + \sum_{\alpha=1}^{N} \left(\mathbf{A}_\alpha - \mathbf{A}\right) \, \Delta \mathrm{H}_\alpha(x-x_\alpha) \, ,
	\label{eq011}
\end{equation}
where 
\begin{equation}
    \Delta \mathrm{H}_\alpha(x) = 
     \mathrm{H} \left( x + \frac{\Delta x_{\alpha}}{2} \right) - \mathrm{H} \! \left( x  - \frac{\Delta x_{\alpha}}{2} \right) 
\end{equation}
stands for the finite step function of width $\Delta x_\alpha$ and centred at the origin,  expressed in terms of the classical Heaviside's unit function \cite{lighthill1964introduction}, $\mathrm{H}(x)$, equal to $0$ for $x < 0$ and $1$ for $x > 0$. Substituting Eq.~\eqref{eq011} into Eq.~\eqref{eq010} yields

\begin{equation}
	\frac{\textrm{d} \mathbf{u}}{\textrm{d} x} = \mathbf{A} \, \mathbf{u} + 
								\sum_{\alpha=1}^{N} \left(\mathbf{A}_\alpha - \mathbf{A}\right) \, \mathbf{u}(x) \,  \Delta \mathrm{H}_\alpha (x-x_\alpha)
								+ \mathbf{f}(x) \, .
	\label{eq056}
\end{equation}
Denoting and defining $\mathbf{f}_r(x,\mathbf{u})$ to be the response of the inclusions, i.e.
\begin{equation}
	\mathbf{f}_r(x,\mathbf{u}) =\sum_{\alpha} \left(\mathbf{A}_\alpha - \mathbf{A}\right) \, \mathbf{u}(x) \,  \Delta \mathrm{H}_\alpha(x-x_\alpha) \, ,
	\label{eq061}
\end{equation}
we consider the solution to
\begin{equation}
	\frac{\textrm{d} \mathbf{u}}{\textrm{d} x} = \mathbf{A} \, \mathbf{u} + 
	\mathbf{f}_r(x,\mathbf{u}) 
	+ \mathbf{f}(x) \, .
	\label{eq057}
\end{equation}
The term $\mathbf{f}_r(x,\mathbf{u})$  can be interpreted as a radiated wavefield in response to an incident wave interacting with the scatterers within the waveguide. The radiated field is solely due to $\mathbf{f}_r(x,\mathbf{u})$, which we express as a series of outgoing sources to account for the contribution of every scatterer and to satisfy the Sommerfeld radiation condition at infinity \cite{sommerfeld1949partial}. Eq.~\eqref{eq010} can be solved analytically using the TMM ~\cite{Rui-2019} depending upon the incident field (e.g. an incoming plane wave) or incident forcing (e.g. a point source). The TMM, despite providing the exact solution of the problem, has the disadvantage of expressing it in the form of a piecewise continuous function; often, as in the case that many inclusions are considered within the waveguide, the numerical implementation utilising piecewise continuous functions from the TMM is unnecessary and not always straightforward. \\

The original Thomson Haskell \cite{thomson1950transmission,haskell1990dispersion} TMM formulation experiences problems with waveguides that feature evanescent modes (e.g. the Euler-Bernoulli beam or higher order rod models), as these  modes create exponentially small or large terms in the underlying matrices from which the solutions are constructed resulting in inaccurate computational results from poorly-conditioned matrices. Methods exist, based upon the seminal work of Dunkin \cite{dunkin1965computation} (see the review article \cite{lowe1995matrix}), to rearrange these matrices to remove the exponentially small or large contributions and hence obtain accurate results from well conditioned matrices. However, the drawback in doing so is that the straightforward solution of the original Thomson Haskell TMM formulation is lost. Other methods to remove numerical instabilities from the problem include constructing well behaved parts of the solution from matrices that are well conditioned, as in the scattering-matrix approach \cite{ko1988scattering}, however this approach is limited to finding only reflection and transmission coefficients to problems. Schemes have been proposed for the full solution based upon numerically stable algorithms, for instance utilising the eigendecomposition of matrices \cite{perez2015relations} to design computations which are not overly sensitive to numerical instabilities, and even combine these with coordinate transformation procedures \cite{chandezon1982multicoated,cotter1995scattering} to remove the numerical instabilities altogether. \\

We propose that, instead of focusing on the exact TMM or scattering-matrix solution, it is advantageous to express the solutions to the multiple scattering problem in a much simpler form; we do this by approximating the exact solution through the Green's function and, owing to the first order matrix form \cite{langley1999wave} we expressed the constitutive equations in, as a matrix of Green's functions. Furthermore, by expressing our matrix of Green's functions in canonical form (eigendecomposition) we can easily distinguish between left and right travelling contributions to the solutions and exponentially oscillating/growing/decaying contributions to the solutions. Therefore, with our proposed method we can readily implement the piecewise nature of the solution. Moreover, since we know exactly which quantities are exponentially growing/decaying, we can rearrange the expressions in a simple manner and use numerically stable operations \cite{dunkin1965computation,perez2015relations,wiltshaw2023analytical} to solve the multiple scattering problem without being hindered by potentially poorly conditioned matrices. \\

The solution of  Eq.~\eqref{eq057} can be expressed as the superposition of an incident field $\bm{\psi}_0(x)$ and a scattered field $\mathbf{u}_s(x)$ as follows
\begin{equation}
	\mathbf{u}(x) = \bm{\psi}_0(x) + \mathbf{u}_s(x) \, ,
	\label{eq058}
\end{equation}
where $\bm{\psi}_0(x)$ and $\mathbf{u}_s(x)$ are solutions to the following problems
\begin{eqnarray}
 	\frac{\textrm{d} \bm{\psi}_0}{\textrm{d} x} &=& \mathbf{A} \, \bm{\psi}_0 
 + \mathbf{f}(x) \, , \quad 
  \label{eq059b} \\
 	\frac{\textrm{d} \mathbf{u}_s}{\textrm{d} x} &=& \mathbf{A} \, \mathbf{u}_s + 
 \mathbf{f}_r(x,\mathbf{u})  \, . 
 \label{eq059}
\end{eqnarray}
\subsection{The incident field} \label{incFe}

Two types of incident field will be considered: (i) a simple plane wave associated with a propagating mode of the waveguide, and (ii) a wavefield excited by a point--force $\mathbf{f}(x) = \mathbf{Q}_0 \, \delta(x-x_0)$ located at certain point $x_0$. \\

For case (i), $\textbf{f}(x) = 0$ in Eq. \eqref{eq059b}, and the incident field can be written as the corresponding mode. For instance, the pair $\{\mathbf{u}_0,  i  k_0\}$ corresponding to the eigenvector and eigenvalue of the matrix $\mathbf{A}$ are associated with a certain propagating mode of the empty-waveguide. Hence, the rightward and leftward incident fields satisfying Eq. \eqref{eq059b} can be written as
\begin{equation}
	\bm{\psi}_0(x) = \mathbf{u}_0^\ast \, e^{-i k_0 x} \ \text{(rightwards)} \ , \qquad
	\bm{\psi}_0(x) = \mathbf{u}_0 \, e^{i k_0 x} \ \text{(leftwards)}	\ ,
	\label{eq063}
\end{equation}
where superscript $\ast$ denotes the complex--conjugate. \\

In case (ii), the incident wave requires the solution of the non--homogeneous Eq.~\eqref{eq059b} for  an excitation term $\mathbf{f}(x) = \mathbf{Q}_0 \, \delta(x-x_0)$, i.e.
\begin{equation}
	\bm{\psi}_0(x) = e^{\mathbf{A}x} \left\lbrace \, \bm{\psi}_0(0) + \int_{0}^{x}  e^{- \mathbf{A}\eta} \, \mathbf{Q}_0 \,  \delta(\eta-x_0) \, d\eta \right\rbrace
	= 
	\begin{cases}
		e^{\mathbf{A}x} \, \bm{\psi}_0(0) 				\quad  &x < x_0 \, , \\
		e^{\mathbf{A}x} \, \bm{\psi}_0(0)   +  e^{\mathbf{A}(x-x_0)} \, \mathbf{Q}_0   	\quad  	\quad  &x > x_0 \, .
	\end{cases}								
	\label{eq064} 
\end{equation}
By definition, the solution in Eq. \eqref{eq064} is the Green's function of the homogeneous waveguide. We express the Green's function in matrix form, involving the Green's functions of each variable of the state--vector. Thus, the solution of the incident field can be expressed as
\begin{equation}
	\bm{\psi}_0(x) = \mathbf{G}(x - x_0) \, \mathbf{Q}_0 \, ,
	\label{eq065}
\end{equation}
where the matrix $\mathbf{G}(x)$ is derived in Sec. \ref{Green} in terms of the eigenvalues and eigenvectors of $\mathbf{A}$.

\subsection{The scattered field}

The scattered field $\mathbf{u}_s(x)$ is given by the solution to Eq. \eqref{eq059}. Assuming $N$ scatterers exist within the waveguide, we decompose $\mathbf{u}_s(x)$ into the superposition of the effects of each scatterer independently yielding the sum 
\begin{equation}
	\mathbf{u}_s(x) = 	\mathbf{u}_s^{(1)}(x)  + \cdots + \mathbf{u}_s^{(N)}(x) \, ,
	\label{eq074}
\end{equation}
where each $\mathbf{u}_s^{(\alpha)}(x)$ is a solution of the differential equation
\begin{equation}
	\frac{\textrm{d} \mathbf{u}_s^{(\alpha)}}{\textrm{d} x} = \mathbf{A} \, \mathbf{u}_s^{(\alpha)} + 
	\left(\mathbf{A}_\alpha - \mathbf{A}\right) \, \mathbf{u}(x) \,  \Delta \mathrm{H}_\alpha(x-x_\alpha) .
	\label{eq075}
\end{equation}
The solution of the above expression is simply
\begin{equation}
			\mathbf{u}_s^{(\alpha)}(x) = e^{\mathbf{A}x} \left[ \mathbf{u}_s^{(\alpha)}(0) + 
		\textbf{I}^{(\alpha)} (x) \right] \, , \quad \quad 
		\label{eq076} 
\end{equation}
where
\begin{equation}
  \textbf{I}^{(\alpha)} (x) = \int_{0}^{x}  e^{-\mathbf{A} \eta} \left(\mathbf{A}_\alpha - \mathbf{A}\right) \, \mathbf{u}(\eta) \,  \Delta \mathrm{H}_\alpha(\eta-x_\alpha) \, d\eta \, .
\end{equation}
Therefore, the solution in Eq. \eqref{eq076}  depends upon the total field $\mathbf{u}(x)$ only inside the inclusion $\alpha$  (as the integrand in $\textbf{I}^{(\alpha)} (x)$ is zero outside of the inclusion), where we assume $\mathbf{u}(x)$ is consistent with the TMM and depends analytically on the state--vector at $x=x_\alpha$, as follows
\begin{equation}
\mathbf{u}(x) = e^{\mathbf{A}_\alpha  (x-x_\alpha)} \, \mathbf{u}(x_\alpha) 
\ , \qquad  \mbox{when} \, \left| x - x_\alpha \right| < \Delta x_\alpha / 2 .
	\label{eq078} 
\end{equation}
Denoting and defining $x_{-}^{(\alpha)}$ for any $x \in [0, \XMfrac)$, it is clear that $\textbf{I}^{(\alpha)}(x_{-}^{(\alpha)}) = 0$ as the integrand is zero. Similarly, denoting $x_{+}^{(\alpha)}$ for any $x \in ( \XPfrac, \infty ]$, we apply the standard properties of Heaviside's unit function and find
\begin{equation}
\begin{split}
    \textbf{I}^{(\alpha)}(x_{+}^{(\alpha)}) = 
    \left( \int_{\XM}^{\XP} e^{-\textbf{A} \eta} \textbf{A}_{\alpha} e^{\textbf{A}_{\alpha} \eta}  \, d \eta \right) &  e^{-\textbf{A}_{\alpha} x_{\alpha}} \textbf{u}(x_{\alpha})  - \\
    - & \left( \int_{\XM}^{\XP} \textbf{A} e^{-\textbf{A} \eta}  e^{\textbf{A}_{\alpha} \eta}  \, d \eta \right) e^{-\textbf{A}_{\alpha} x_{\alpha}} \textbf{u}(x_{\alpha}) \, . 
    \end{split} \label{EqIalpha}
\end{equation}
From Eq. ~\eqref{EqIalpha}, by integrating the first integral by parts one immediately finds   
\begin{equation}
    \textbf{I}^{(\alpha)}(x_{+}^{(\alpha)}) = 
    \left[ e^{- \mathbf{A} \, (\XP)} \ e^{\mathbf{A}_\alpha  \, (\frac{\Delta x_{\alpha}}{2} )} - 
											 e^{-\mathbf{A} \, (\XM) } \ e^{-\mathbf{A}_\alpha  \, (\frac{\Delta x_{\alpha}}{2})} \right] \textbf{u}(x_{\alpha}) \, ,
\end{equation}
and hence that
\begin{equation}
    \textbf{I}^{(\alpha)}(x) = \begin{cases}
        0  		\quad   & x < \XM \, , \\
        \left[ e^{- \mathbf{A} \, (\XP)} \ e^{\mathbf{A}_\alpha  \, (\frac{\Delta x_{\alpha}}{2} )} - 
											 e^{-\mathbf{A} \, (\XM) } \ e^{-\mathbf{A}_\alpha  \, (\frac{\Delta x_{\alpha}}{2})} \right] \textbf{u}(x_{\alpha})  	\quad   & x > \XP \, .
    \end{cases} \label{Ialpha+-} 
\end{equation}
Substituting Eq. ~\eqref{Ialpha+-} into Eq. ~\eqref{eq076} one finds
\begin{equation}
	\mathbf{u}_s^{(\alpha)}(x) 
	= 
	\begin{cases}
		e^{\mathbf{A}x} \, \mathbf{u}_s^{(\alpha)} (0) 				\quad  & x < x_\alpha - \Delta x_\alpha/2, \\
		e^{\mathbf{A}x} \, \mathbf{u}_s^{(\alpha)}(0)   +  
		e^{\mathbf{A} (x - x_\alpha)} \, \mathbf{Q}_\alpha   	\quad  	\quad  & x > x_\alpha + \Delta x_\alpha/2,
	\end{cases}								
	\label{eq081} 
\end{equation}
where 
\begin{align}
    \textbf{Q}_{\alpha} & = \textbf{K}_{\alpha} \textbf{u}(x_{\alpha}) \label{eq086a} \, , \\
    \mathbf{K}_\alpha & = e^{- \mathbf{A} \, \Delta x_\alpha /2 } \ e^{\mathbf{A}_\alpha  \, \Delta x_\alpha /2} - 
											 e^{\mathbf{A} \, \Delta x_\alpha /2 } \ e^{-\mathbf{A}_\alpha  \, \Delta x_\alpha /2} \, .
											 \label{eq086}
\end{align}
Note that when the host material and inclusion properties match $\mathbf{K}_\alpha = \textbf{0}$, and as expected no scattering  occurs in this case. In section \ref{Green}, we make use of the eigendecomposition of the exponential of a matrix in terms of the eigenvalues and eigenvectors of their arguments, i.e.
\begin{equation}
	e^{\mathbf{A}x} = \sum_{j=1}^{2m} \, \mathbf{u}_j \, \mathbf{v}_j^T \,  	e^{{\lambda}_j \, x} \ , \quad	
	e^{\mathbf{A}_\alpha x} = \sum_{j=1}^{2m} \, \mathbf{x}_j \, \mathbf{y}_j^T \,  	e^{{\lambda}_j^{(\alpha)} \, x} \, ,
		\label{eqB05}
\end{equation}
where superscript $T$ denotes the transpose of a vector. The triple $\mathbf{u}_j, \ \mathbf{v}_j$ and $\lambda_j$ correspond to the $j$th right and left eigenvector and eigenvalue of matrix $\mathbf{A}$, which satisfy the following eigenrelations 
\begin{equation}
	\mathbf{A} \, \mathbf{u}_j = \lambda_j \, \mathbf{u}_j \ , \quad 
	\mathbf{v}_j^T \, \mathbf{A}  = \lambda_j \, \mathbf{v}_j^T \ , \quad 	
	\mathbf{v}_j^T \, \mathbf{u}_l  = \delta_{jl} \quad 1 \leq j,l \leq 2m \, .
	\label{eqB06}
\end{equation}
Here, $\delta_{jl}$ is the Kronecker delta function. Similar definitions apply to $\mathbf{x}_j, \ \mathbf{y}_j$ and $\lambda_j^{(\alpha)}$ for the matrix $\mathbf{A}_{\alpha}$. \\

For asymptotically small inclusions, the solution can be directly expressed in terms of the Green's function. Indeed since, by treating $\Delta x_{\alpha}$ as a small parameter and considering Eq. \eqref{eq075} in the limit as $\Delta x_{\alpha} \to 0$, by the formal definition of the derivative we find
\begin{equation}
	\lim_{\Delta x_{\alpha} \to 0} \left\lbrace \frac{\textrm{d} \mathbf{u}_s^{(\alpha)}}{\textrm{d} x} - \mathbf{A} \, \mathbf{u}_s^{(\alpha)} \right\rbrace =
	\left(\mathbf{A}_\alpha - \mathbf{A}\right) \, \mathbf{u}(x) \, \Delta x_{\alpha} \,   \frac{d}{dx} \Big\lbrace \mathrm{H} ( x - x_{\alpha} ) \Big\rbrace = \left(\mathbf{A}_\alpha - \mathbf{A}\right) \, \mathbf{u}(x) \, \Delta x_{\alpha} \,   \delta ( x - x_{\alpha} ) .
	\label{eq075dash}
\end{equation}
Here, we have exploited the well known fact that the derivative of Heaviside's unit function is the Dirac delta function \cite{lighthill1964introduction}, i.e.
\begin{equation}
    \delta(x) = \lim_{\Delta x_{\alpha} \to 0} \frac{\Delta \mathrm{H}_\alpha(x)}{\Delta x_{\alpha} } = 
    \frac{d}{dx} \Big\lbrace \mathrm{H} ( x ) \Big\rbrace \, .
    \label{eq_heaviside}
\end{equation}

The solution of Eq. \eqref{eq075dash} is given in Eq. \eqref{eq076}, where now $\textbf{I}^{(\alpha)} (x)$ is given by
\begin{equation}
  \textbf{I}^{(\alpha)} (x) = \int_{0}^{x}  e^{-\mathbf{A} \eta} \left(\mathbf{A}_\alpha - \mathbf{A}\right) \, \mathbf{u}(\eta) \, \Delta x_{\alpha}  \, \delta(\eta - x_{\alpha}) \, d\eta \, .
\end{equation}
Therefore, the analogue of $ \textbf{u}_{s}^{(\alpha)}$ in Eq. \eqref{eq081} for asymptotically small inclusions is
\begin{equation}
    \textbf{u}_{s}^{(\alpha)} = \begin{cases}
        e^{\mathbf{A}x} \, \mathbf{u}_s^{(\alpha)} (0) 		&		\quad  x < x_\alpha \\
        e^{\mathbf{A}x} \, \mathbf{u}_s^{(\alpha)}(0)   +  
		e^{\mathbf{A} (x - x_\alpha)} \, \widetilde{\textbf{Q}}_{\alpha} 		&		\quad  x > x_\alpha  
    \end{cases} \quad \quad \mbox{as $\Delta x_{\alpha} \to 0$,} \label{eq081SMALL}
\end{equation}
where
\begin{align}
    \widetilde{\textbf{Q}}_{\alpha} & = \widetilde{\textbf{K}}_{\alpha} \textbf{u}(x_{\alpha}) \label{eq086Dasha}, \\
    \widetilde{\textbf{K}}_{\alpha} & = (\textbf{A}_{\alpha} - \textbf{A}) \Delta x_{\alpha}. \label{eq086Dash}
\end{align}

Establishing an analogy between the general solution given in Eqs.~\eqref{eq081}-\eqref{eq086} and that given for asymptotically small inclusions Eqs.~\eqref{eq081SMALL}-\eqref{eq086Dash}, it is clear that the terms of the vector $\mathbf{Q}_\alpha$ physically behave as source terms that radiate waves outward proportional to the excitation of the internal degrees of freedom $\mathbf{u}(x_\alpha)$. These forces are distributed inside the scatterer over a finite but small length $\Delta x_\alpha$. We hypothesize that, even in the case that $\Delta x_{\alpha}$ is not asymptotically small, such radiation will be well approximated as if it were produced by a point force placed at $ x_\alpha$ within the empty-waveguide. We expect that this approximation will be accurate provided that the wavelength considered is much larger than $\Delta x_{\alpha}$. In section \ref{NES1} and \ref{NES2} we provide test cases for the approximate solution by cross--checking against the exact TMM solution, where we define a parameter to estimate the upper bound of the relative error introduced by this approximation. For now, we assume that Eq.~\eqref{eq081} will be well approximated by
\begin{equation}
	\mathbf{u}_s^{(\alpha)}(x) 
	\approx 
	\begin{cases}
		e^{\mathbf{A}x} \, \mathbf{u}_s^{(\alpha)} (0) 		&		\quad  x < x_\alpha \, , \\
		e^{\mathbf{A}x} \, \mathbf{u}_s^{(\alpha)}(0)   +  
		e^{\mathbf{A} (x - x_\alpha)} \, \mathbf{Q}_\alpha  & 	\quad  	\quad  x \geq x_\alpha \, .
	\end{cases}								
	\label{eq082} 
\end{equation}

Subsequently, it is convenient to express the solutions in \eqref{eq081SMALL} and \eqref{eq082} in terms of the Green's function of the homogeneous waveguide, as in Eq. \eqref{eq065}, as follows
\begin{equation}
	\mathbf{u}_s^{(\alpha)}(x)  = \mathbf{G}(x - x_\alpha ) \, \mathbf{K}_\alpha \mathbf{u}(x_\alpha) \, .
	\label{eq083}
\end{equation}
The total scattered field within the waveguide is $	\mathbf{u}_s(x) = \sum_\alpha 	\mathbf{u}_s^{(\alpha)}(x)  $, yielding
\begin{equation}
	\mathbf{u}_s(x)  = \sum_{\alpha = 1}^{N}\mathbf{G}(x - x_\alpha ) \, \mathbf{K}_\alpha \mathbf{u}(x_\alpha) \, .
	\label{eq084}
\end{equation}
Hence, small scatterers within the waveguide can be thought of as a regular perturbation to the empty-waveguide, such a perturbation being a monopole point--force term applied to the empty-waveguide at the centre of every approximated scatterer. This method is particularly useful for one-dimensional waveguides, as the Green's functions are regular and one need not concern themselves with the difficulties associated with singular perturbation problems \cite{schnitzer2017bloch,Wiltshaw-2020,wiltshaw2023analytical}. Thus, the equivalent approach proposed in this paper to determine the solution of the multiple scattering problem is summarized in the following differential equation
\begin{equation}
	\frac{\textrm{d} \mathbf{u}}{\textrm{d} x} = \mathbf{A} \, \mathbf{u} + \sum_{\alpha = 1}^{N} \mathbf{K}_\alpha \mathbf{u}(x_\alpha) \delta(x - x_\alpha) + \mathbf{f}(x) \, .
	\label{eq084b}
\end{equation}
In this expression, $\mathbf{u}(x_\alpha)$ is the state--vector evaluated at the centre of the scatterers and at this stage it is unknown. The components of $\mathbf{u}(x_\alpha)$ will be determined by solving a linear system of equations which will be introduced later. The matrix $\mathbf{K}_\alpha$ is given in Eq. \eqref{eq086} and depends on the scatterer size $\Delta x_\alpha$ and on the contrast between elastodynamic properties of the empty-waveguide and of the scatterers. Therefore, $\mathbf{K}_\alpha \mathbf{u}(x_\alpha)$ approximates a scatterer of finite width as the coefficient of a point-source term. In this regard, it is interesting to consider if this approximation reveals anything about the physical process of scattering. For instance, consider the simple model for longitudinal waves and the meaning of the vector $\mathbf{Q}_\alpha = \mathbf{K}_\alpha \mathbf{u}(x_\alpha)$. As known from Table~\ref{tab01}, the state--vector for longitudinal waves is $\mathbf{u}(x) = \{u(x),N_x(x)\}$, whose components are the axial displacement and normal force. Assuming that the inclusion width is much smaller than the wavelength, i.e. $k \Delta x_\alpha \ll 1$, we can expand the exponential matrices of Eq.~\eqref{eq086} as
\begin{equation}
e^{\pm \mathbf{A} \Delta x_\alpha \eta } = \mathbf{I} \pm \mathbf{A} \Delta x_\alpha \eta + \mathcal{O}(\Delta x_\alpha^2) \ , \qquad
e^{\pm \mathbf{A}_\alpha \Delta x_\alpha \eta } = \mathbf{I} \pm \mathbf{A}_\alpha \Delta x_\alpha \eta + \mathcal{O}(\Delta x_\alpha^2), \quad \quad \mbox{as $\Dxa \to 0$}. \ 
\label{eq085}
\end{equation}
This leads to the first order approximation of $\mathbf{K}_\alpha$ in terms of $\Delta x_\alpha$, hence from Eq.~\eqref{eq086Dash}
\begin{equation}
	\mathbf{K}_\alpha = \widetilde{\mathbf{K}}_\alpha + \mathcal{O}(\Delta x_\alpha^2) \sim 
	\begin{bmatrix}
		0 & 1/EA_\alpha - 1/EA \\
		(\rho A - \rho A_\alpha) \omega^2 & 0
	\end{bmatrix} \, \Delta x_\alpha, \quad \mbox{as $\Delta x_\alpha \to 0$}.
\end{equation}
Therefore, Eq. \eqref{eq086a} yields 
\begin{equation}
	\mathbf{Q}_\alpha \sim
	\left\{
	\begin{array}{c}
	\left(\frac{\Delta x_\alpha}{EA_\alpha} -  	\frac{\Delta x_\alpha}{EA}\right) \, N_x(x_\alpha)\\
	\left( \rho A - \rho A_\alpha\right) \Delta x_\alpha \omega^2  \, u(x_\alpha)
	\end{array}\right\} \, , \quad \mbox{as $\Delta x_\alpha \to 0$}.
\end{equation}
The magnitudes $K_\alpha = EA_\alpha / \Delta x_\alpha$  and $K = EA / \Delta x_\alpha$ are the linear rigidities of two springs corresponding to the elastic longitudinal behaviour of a rod--segment $\Delta x_\alpha$, for both the inclusion and the surrounding medium respectively. Hence, the value $\Delta u = N_x(x_\alpha)/K_\alpha - N_x(x_\alpha)/K$ can be interpreted as a small jump in the displacement fields due to the presence of an inhomogeneity at $x=x_\alpha$.The second term represents the inertia force (mass per acceleration) produced in a segment $\Delta x_\alpha$ due to the difference in masses, $ m = \rho A \Delta x_\alpha$ and $ m _\alpha= \rho A_\alpha \Delta x_\alpha$. This difference $\Delta N_x = - \left( m_\alpha- m \right)  \omega^2  \, u(x_\alpha)$ represents a new inertia force which will propagate along the rod via the Green's function as a scattering radiation, as shown in Eq.~\eqref{eq083}. The scattering effect depends strongly on both the width of the inclusion and on the contrast between properties. The terms of vector $\mathbf{Q}_\alpha$ for longitudinal waves can be read as
\begin{equation}
	\mathbf{Q}_\alpha =
	\left\{
	\begin{array}{c}
		\left(\frac{1}{K_\alpha} -  	\frac{1}{K}\right) \, N_x(x_\alpha)\\
		\left(  m -  m_\alpha \right)  \omega^2  \, u(x_\alpha)
	\end{array}\right\} \equiv
	\left\{
\begin{array}{c}
	\Delta u\\
	\Delta N_x
\end{array}\right\} \, .
\end{equation}
Similar interpretations can be made for other types of waveguides. For instance, considering a Timoshenko beam (see Table \ref{tab03}), the vector $\mathbf{Q}_\alpha$ after the same simplification becomes
\begin{equation}
	\mathbf{Q}_\alpha \sim 
	\left\{
	\begin{array}{c}
	\left(\frac{\Delta x_\alpha}{GA_\alpha} -  	\frac{\Delta x_\alpha}{GA}\right) \, V_z(x_\alpha) \\
	\left(\frac{\Delta x_\alpha}{EI_\alpha} -  	\frac{\Delta x_\alpha}{EI}\right) \, M_y(x_\alpha) \\ 	
	- \left( \rho A_\alpha- \rho A\right) \Delta x_\alpha \omega^2  \, w(x_\alpha) \\ 
	- \left( \rho I_\alpha - \rho I\right) \Delta x_\alpha \omega^2  \, \varphi_y(x_\alpha)	
	\end{array}
	\right\}
	\equiv
		\left\{
	\begin{array}{c}
		\Delta w\\
		\Delta \varphi_y \\
		\Delta V_z \\
		\Delta M_y
	\end{array}\right\} \, , \quad \mbox{as $\Delta x_\alpha \to 0$}.
\end{equation}
Here, the two first components of $\mathbf{Q}_\alpha$ represent a displacement due to the shear strain and a rotation due to the bending flexibility respectively. Similarly, the last two entries represent inertia forces produced by the difference in mass and rotational inertia of the cross section.

\section{Solution of the multiple scattering problem}

The solution of the total field involves determining the state--vector at $x=x_\alpha$, and hence the scattering problem compatibility-equations must be established. As introduced above, let us consider an incident field $\bm{\psi}_0(x)$ and the scattered field $\mathbf{u}_s(x)$ radiated from the inclusions. From Eq.~\eqref{eq084}  the total field is then 
\begin{equation}
	\mathbf{u}(x) = \bm{\psi}_0(x) + \mathbf{u}_s(x) = \bm{\psi}_0(x) + \sum_{\beta=1}^N \mathbf{G}(x-x_\beta) \, \mathbf{K}_\beta \, \mathbf{u}(x_\beta) \, ,
	\label{eq013}
\end{equation} 
where $\mathbf{G}(x)$ is the matrix of  Green's functions of the system (matrix of the same size as $\mathbf{A}$) which will be deduced in Sec. \ref{Green}. Above, the variables $\mathbf{u}(x_\alpha)$ are unknown for every $\alpha = 1, 2, \ldots, N$, their solution is found after evaluating Eq.~\eqref{eq013} at $x = x_1, \ldots, x_N$ leading to the following system of linear equations
\begin{equation}
\mathbf{u}(x_\alpha) - \sum_{\beta=1}^N \mathbf{G}(x_\alpha-x_\beta) \, \mathbf{K}_\beta \, \mathbf{u}(x_\beta)= \bm{\psi}_{0}(x_\alpha)
\, , \quad \ \alpha = 1, 2, \ldots, N  \, .
\label{eq014} 
\end{equation}
Since the size of vector $\mathbf{u}(x)$ is $2m$, then Eq. \eqref{eq014} represents a system of $2mN$ equations for $2mN$ unknowns.  After some rearrangements, Eq.~\eqref{eq014} can be written as
\begin{equation}
	\left[
	\begin{array}{cccc}
		\mathbf{I}	- \mathbf{G}(0^+) \mathbf{K}_1			&			- \mathbf{G}(x_1-x_2) \mathbf{K}_2   & \cdots 	&- \mathbf{G}(x_1-x_N) \mathbf{K}_N \\
		- \mathbf{G}(x_2-x_1) \mathbf{K}_1  & \mathbf{I}	- \mathbf{G}(0^+) \mathbf{K}_2		   & \cdots 	&- \mathbf{G}(x_2-x_N) \mathbf{K}_N \\
		\vdots					 & 				\vdots															 &   \ddots  &   \vdots  \\
		- \mathbf{G}(x_N-x_1) \mathbf{K}_1			&	- \mathbf{G}(x_N-x_2) \mathbf{K}_2   & \cdots 	& 	\mathbf{I}	- \mathbf{G}(0^+) \mathbf{K}_N
	\end{array}
	\right]
	\left\{
	\begin{array}{c}
		\mathbf{u}(x_1) \\ 		\mathbf{u}(x_2)  \\ \vdots \\ 		\mathbf{u}(x_N)
	\end{array}\right\}	
	=
	\left\{
	\begin{array}{c}
		\bm{\psi}_0(x_1) \\ 		\bm{\psi}_0(x_2) \\ \vdots \\ 		\bm{\psi}_0(x_N)
	\end{array}\right\} \, ,
	\label{eq043}
\end{equation} 
with $N$ unknown state--vectors on the left hand side, each of dimension $2m$, hence the $2mN$ unknowns are given by all of the components of 
\begin{equation}
		\mathbf{u}(x_1) \ , \ldots , \mathbf{u}(x_N) \, .
		\label{eq015}
\end{equation}
See, for example Table \ref{tab01} and \ref{tab03} for the explicit unknown components within $\mathbf{u}(x_\alpha)$ which need to be determined for different models of the waveguide.

\section{Closed-form derivation of the Green's matrix}
\label{Green}

The procedure developed herein makes it possible to treat one-dimensional structures with a distribution of $N$ inclusions as if it were a homogeneous medium under the influence of $N$ point sources. Since the derivations are valid for any homogeneous waveguide satisfying Eq. \eqref{eq001}, we develop  the Green's matrix $\mathbf{G}(x)$ in terms of the matrix $\textbf{A}$ so that the expressions can be applied to simulate multiple scattering problems for any such waveguide.\\

A waveguide governed by the matrix $\mathbf{A}$ with state--vector $\mathbf{u}$ will present in general $2m$ modes, with $m$ modes corresponding to rightwards waves and $m$ modes corresponding to leftwards waves. These modes can be either propagating or evanescent depending on the nature of the model considered. For instance, longitudinal waves in rods have $m=1$ propagating modes in each direction. Low frequency flexural waves present $2m=4$ modes, one propagating and one evanescent for both rightward and leftward waves. High frequency beam waves (Timoshenko beam) present $m=2$ propagating modes at each direction (bending and shear waves). Denoting $\mathbf{u}_j$ and $\mathbf{v}_j$ to be the right and left eigenvectors associated to each mode with eigenvalue $\lambda_j$, i.e. for the matrix $\textbf{A}$ governing the conserved quantities along the empty-waveguide
\begin{equation}
	\mathbf{A} \mathbf{u}_j = \lambda_j \, \mathbf{u}_j \ , \quad 
	\mathbf{v}_j^T \, \mathbf{A}  = \lambda_j \, \mathbf{v}_j^T \ , \quad 	
	\mathbf{v}_j^T \, \mathbf{u}_l  = \delta_{jl} \, , \quad 1 \leq j,l \leq 2m
	\label{eq066}
\end{equation}
where $\delta_{jl} $ denotes the Kronecker delta function. Consider the wavefield due to the point--force excitation  $\mathbf{f}(x) = \mathbf{Q}(x_0) \, \delta(x-x_0)$,  at any point along the waveguide by Eq.~\eqref{eq064} we have
\begin{equation}
	\mathbf{u}(x) = 
	\begin{cases}
		e^{\mathbf{A}x} \, \mathbf{u}(0) 		&		\quad  x < x_0 \, ,\\
		e^{\mathbf{A}x} \, \mathbf{u}(0)   +  e^{\mathbf{A}(x-x_0)} \, \mathbf{Q}(x_0)  & 	\quad  	\quad  x > x_0 \, .
	\end{cases}								
	\label{eq067} 
\end{equation}
Moreover, since we are dealing with infinite waveguides, the wavefield solution radiates waves in both directions, rightwards for $x>x_0$ and leftwards for $x<x_0$. Therefore, ordering our eigenvalues such that the first $m$ modes represent rightward waves and the last $m$ leftward waves, the solution may be expressed as follows
\begin{equation}
	\mathbf{u}(x) = 
	\begin{cases}
		\phantom{-}\sum_{j \leq m} C_j \, \mathbf{u}_j \, e^{\lambda_j x}  & \quad x > x_0 \, , \\
		-\sum_{j > m} C_j \, \mathbf{u}_j \, e^{\lambda_j x}  & \quad x < x_0		\, ,
		\label{eq069}						
	\end{cases}
\end{equation}
where the coefficients $C_j, \ 1 \leq j \leq 2m$ need to be determined. In the above,  the negative sign is a formal convention. Evaluating Eq. \eqref{eq069} at $x=x_0^-$ and at $x=x_0^+$ we have
\begin{align}
	e^{\mathbf{A}x_0} \, \mathbf{u}(x_0) 	 						   & = -\sum_{j > m} C_j \, \mathbf{u}_j \, e^{\lambda_j x_0} \ , \qquad (x_0=x_0^-) \label{eq068a}\\
	e^{\mathbf{A}x_0} \, \mathbf{u}(x_0)  + \mathbf{Q}(x_0)	 & = \phantom{-}\sum_{j \leq m} C_j \, \mathbf{u}_j \, e^{\lambda_j x_0}   \ , \qquad (x_0=x_0^+) \, .
	\label{eq068}
\end{align}
The above equations represent a system of $4m$ linear equations with $4m$ unknowns from which we can obtain: $2m$ components of $\mathbf{u}(0)$ and the $2m$ coefficients $C_j, \ 1 \leq j \leq 2m$. Subtracting Eq. \eqref{eq068a} from Eq.~\eqref{eq068} we find
\begin{equation}
	\sum_{j =1 }^{2m} C_j \, \mathbf{u}_j \, e^{\lambda_j x_0} = \mathbf{Q}	(x_0) \, .
	\label{eq070}
\end{equation}
Premultiplying Eq. \eqref{eq070} by $\mathbf{v}_l^T$ and using the orthogonal relationships, we can obtain the explicit value of each coefficient $C_l$ as
\begin{equation}
	C_l  = \mathbf{v}_l^T \mathbf{Q}(x_0) \, e^{-\lambda_j x_0} 	\quad , \quad 1 \leq l \leq 2m \, .
	\label{eq071}
\end{equation}
Now, using the above result in Eq.~\eqref{eq069}, we find the following explicit expression for $\mathbf{u}(x)$
\begin{equation}
	\mathbf{u}(x) = \mathbf{G}(x - x_0) \, \mathbf{Q}(x_0) \, ,
	\label{eq072}
\end{equation}
where the matrix of Green's functions $\mathbf{G}(x)$ is finally 
\begin{equation}
	\mathbf{G}(x) = 
	\begin{cases}
		\displaystyle \phantom{-}\sum_{j \leq m}  \mathbf{u}_j  \, \mathbf{v}_j^T  \, e^{\lambda_j x}  & \quad x > 0 \, , \\
		\displaystyle 	-\sum_{j > m}  \, \mathbf{u}_j  \, \mathbf{v}_j^T   \, e^{\lambda_j x}  & \quad x < 0 \, .
	\end{cases}
	\label{eq073}						
\end{equation}
Using the orthogonality conditions, it is straightforward to obtain the following properties of $\textbf{G}(x )$
\begin{align}
	 \mathbf{G}'(x ) & = \mathbf{A} \, \mathbf{G}(x ) \, , \label{P111}\\
	 \mathbf{G}({0}^+) & - \mathbf{G}({0}^- ) = \mathbf{I}_{2m} \, . \label{P112}
\end{align}
Here $\mathbf{I}_{2m}$ denotes the identity matrix of dimension $(2m) \times (2m)$. The property in Eq. \eqref{P112} characterises the jump discontinuity in the state--vector about the origin, as required for the diagonal terms in the matrix of Eq. \eqref{eq043}. Additionally, expressions \eqref{eq073}-\eqref{P112} hold for any point--force source terms from Eq. \eqref{eq013}; therefore, in Eqs. \eqref{eq073}-\eqref{P112} we may evaluate at particular values of the form $x = x_\alpha - x_\beta$, to compute the required blocks of the matrix in Eq. \eqref{eq043}, where $x_\alpha, \ x_\beta$ represent the coordinates of two arbitrary scatterers. The matrix in Eq. \eqref{eq043} is readily invertible, and hence given a known source term $\boldsymbol{\psi}_{0}$, the required state vectors $\textbf{u}(x_{\alpha})$ are readily determined. Once every $\textbf{u}(x_{\alpha})$ is known, the multiple scattering problem is solved as the state--vector field is easily computed over the entire waveguide from Eq. \eqref{eq013}. From here, one can extract the displacement field for instance. To illustrate how simple it is to consider different models within these expressions, we provide several cases of structural 1D models and their respective state vectors - refer to Table \ref{tab03}, where one simply needs to insert the appropriate $\textbf{A}$ into these expressions to solve the multiple scattering problem for any elastic-waveguide of interest.

\section{Numerical example: multiple scattering from classical longitudinal waves in a rod} \label{NES1}

Let us consider an aluminium rod with a 5$\times$5 cm$^2$ square cross--section. The sectional stiffness and mass per unit length of the rod was taken to be $EA = 1.75 \times 10^ 5$ kN/m and $\rho A=$5.25 kg/m respectively. In sections \ref{NES1} and \ref{NES2}, our main objective is to validate the proposed model, checking the range of validity of the point--force approximation for small inclusions in different models. To this end, we investigate how the solution behaves as several parameters vary. Regarding the characteristics of the set of inclusions, we can distinguish between the sectional stiffness and the mass per unit length, denoted by $EA_\alpha$ and $\rho A$ respectively, the width of the inclusion $\Delta x_\alpha$ and the total number of scatterers $N$. \\

\begin{figure}[H]%

\hspace*{-0.45cm}
\scalebox{0.9}{

		\begin{tabular}{cc}
			\includegraphics[width=7cm]{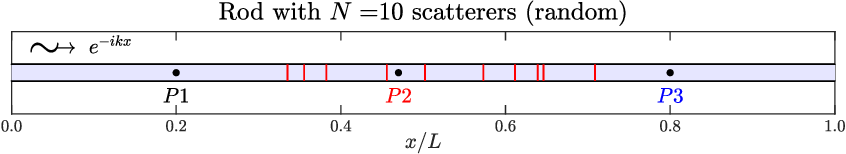} 	&
			\includegraphics[width=7cm]{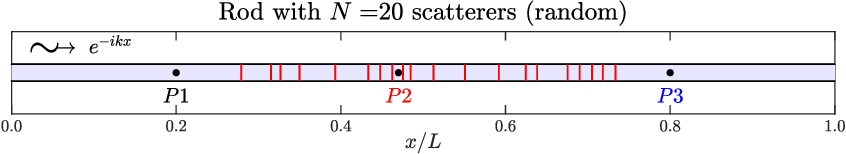}	\\	\\				
			\includegraphics[width=8cm]{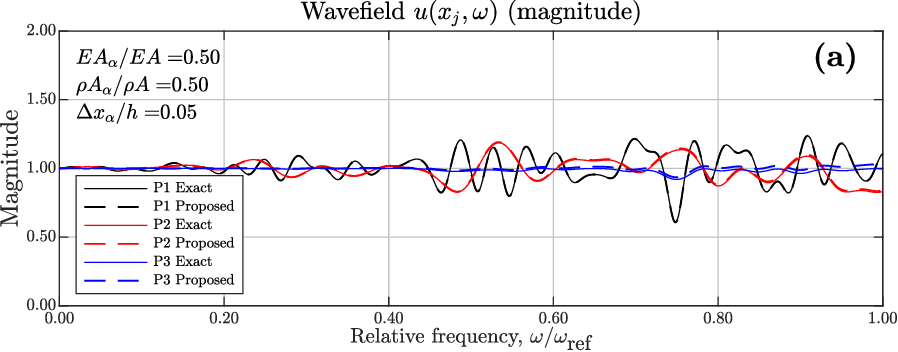} &
			\includegraphics[width=8cm]{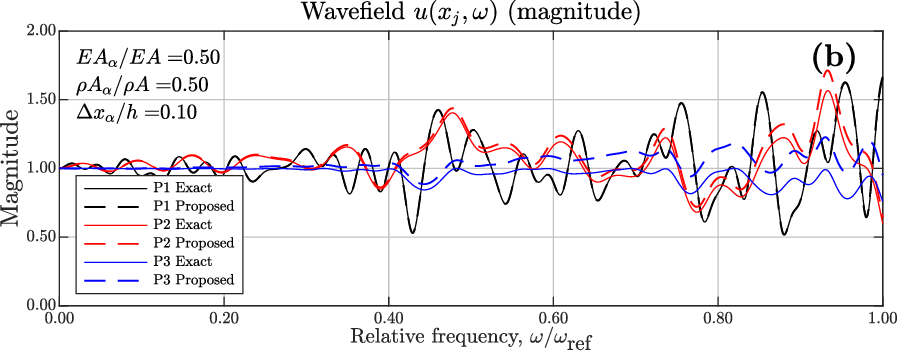} \\ \\		 					    
			\includegraphics[width=8cm]{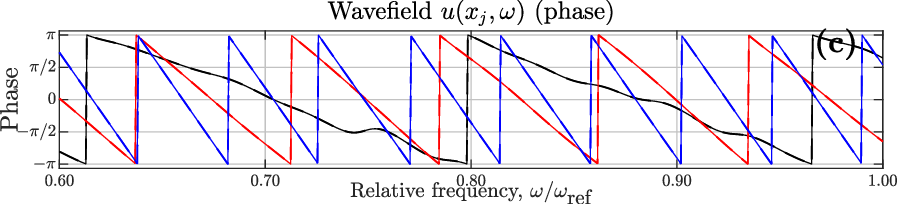} &
			\includegraphics[width=8cm]{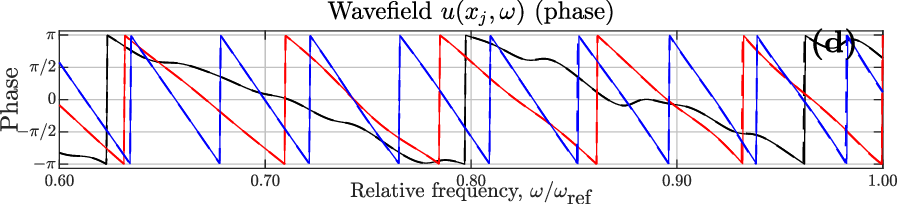} \\ \\		 					    
			\includegraphics[width=8cm]{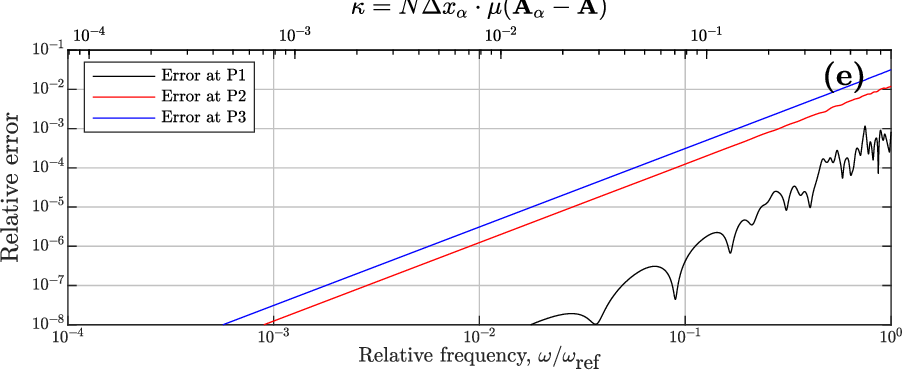} &										   
			\includegraphics[width=8cm]{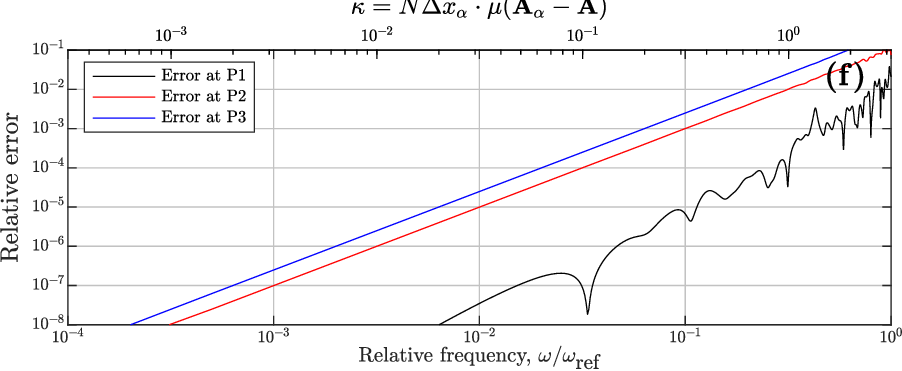}					   
		\end{tabular}
}		
		\caption{Wavefield results of magnitude, phase and error for an infinitely long rod, containing $N=10$ and $\Delta x_\alpha = 0.05h$ (top left) and $N=20$ and $\Delta x_\alpha = 0.1h$  (top right) randomly distributed scatterers, under rightwards incident plane wave excitation, where $h=5$ cm denotes the high of the cross section and $L=4$ m is the length of the region under study. We consider the wavefield at positions $P_1$, $P_2$ and $P_3$,  as defined in the schematic in the top panels where $L = 4 \mathrm{m}$. The relative parameters for both cases here were $EA_\alpha/EA=0.5$ and $\rho A_\alpha / \rho A = 0.5$. Here, we plot the: magnitude of wavefield in (a) and (b), phase of wavefield in (c) and (d), and the  relative error as function of frequency and of the parameter $\kappa$ in (e) and (f).}%
		\label{fig02}%

\end{figure}

In Fig.~\ref{fig02}, the wavefield $u(x,\omega)$ along a $L=4$ m section of an infinitely long rod has been calculated using both the exact solution based on the TMM, and the proposed approach as outlined in the paragraph above section \ref{NES1}. This figure considers the inclusions to be softer than that of the empty-waveguide ($EA_\alpha/EA=0.6$ and $\rho A_\alpha / \rho A = 0.6$) and considers two distributions of scatterers for the following cases:
\begin{itemize}
	\item [(i)] $N=10$ randomly distributed scatterers located within a two meter portion of the rod, with $\Delta x_\alpha = 0.05h = 2.5$ mm  (Figs.~\ref{fig02}(left) ),
	\item [(ii)] $N=20$ randomly distributed scatterers within the same two meter portion of the rod and with $\Delta x_\alpha = 0.10h = 5.0$ mm  (Figs.~\ref{fig02}(right)). 
\end{itemize}
The range of validity of the point-source approximation will be studied within a frequency range consistent with the validity of the model used for the waveguide. The reference frequency, $\omega_{\text{ref}}$, specifying the upper bound in frequency for which the rod model remains accurate is given in Graff \cite{Graff-1999} as 
\begin{equation}
	\omega_{\text{ref}} = 0.30 \sqrt{\frac{EA}{\rho I_O \nu^2}} \approx 40 \ \textnormal{kHz} \, , \label{OmReference}
\end{equation}

\noindent where $I_O$ is the polar moment of inertia of the cross section and $\nu$ is the Poisson coefficient. The limit in Eq.~\eqref{OmReference} is obtained by comparing the classical rod model with the higher-order Love rod-model, where it is established that for $\omega \leq \omega_{\text{ref}}$ there exists a good agreement between both models. Hence,  we shall consider $\omega / \omega_{\text{ref}} \leq 1$ for the frequency range of interest for the comparison between the TMM solution and our approximate solution constructed from the Green's function. \\

The results shown in Figs. \ref{fig02}(left) and Fig. \ref{fig02}(right) consider the wavefield $u(x,\omega) = \left| u \right| e^{i \phi}$, where we plot both the magnitude $\left| u \right| $ and phase $\phi$ of the wave for the cases (i) and (ii) as outlined above.  Comparing $\left| u \right| $ in Figs. \ref{fig02}(a) and Fig. \ref{fig02}(b), we observe that the approximation is valid over the frequency range of interest as long as $N$ or $\Delta x_{\alpha}$ does not get too large. For larger values of $N$ and $\Delta x_\alpha$, especially for higher frequencies where the wavelength becomes smaller, the accuracy of the solution drops off. However, note that the magnitude of the approximate solution is much more affected than its phase, which closely matches the exact solution in the whole frequency range considered, as shown in Fig. \ref{fig02}(c) and Fig. \ref{fig02}(d).\\

According to the theoretical results, the accuracy of the proposed model strongly depends on several physical parameters. These are: the frequency (i.e. the wavelength),  the number of inclusions $N$, their size $\Delta x_\alpha$, and finally the contrast between material parameters represented by $\mathbf{A}_\alpha - \mathbf{A}$. In order to simply study the influence of all of these parameters on the error of the approximation, we introduce the following scattering parameter $\kappa$ defined as
 \begin{equation}
 	\kappa = N \, \Delta x_\alpha \cdot \mu \left(\mathbf{A}_\alpha - \mathbf{A}\right) \, ,
 	\label{eq088}
 \end{equation}
 and we hypothesize that the error introduced by approximating the inclusions as point-source terms can be estimated by $\kappa$. In the above, $\mu(\mathbf{M})$ represents the spectral radius of the matrix $\mathbf{M}$, defined for any diagonalizable matrix $\mathbf{M}$ as \cite{Householder-1964}
 \begin{equation}
 \mu(\mathbf{M}) = \max \left\{\left|\lambda \right| : \ \mbox{$\lambda$ is eigenvalue of $\mathbf{M}$} \right\} \, .
 \end{equation}
 
  \begin{figure}[H]%
\hspace*{-0.45cm}
\scalebox{0.9}{
		\begin{tabular}{cc}
			\multicolumn{2}{c}{\includegraphics[width=8cm]{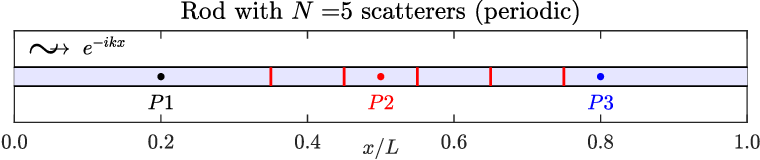}} \\
			\includegraphics[width=8cm]{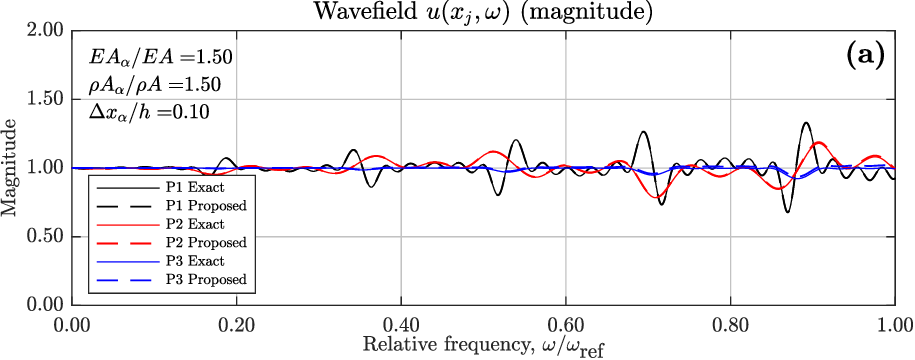} 	&
			\includegraphics[width=8cm]{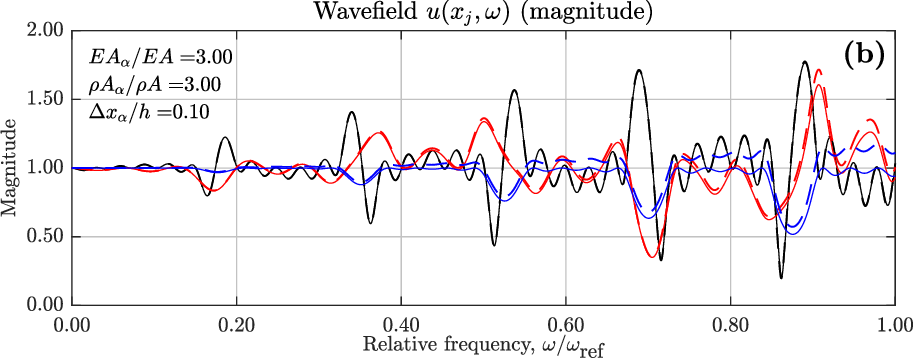} \\ \\			
			\includegraphics[width=8cm]{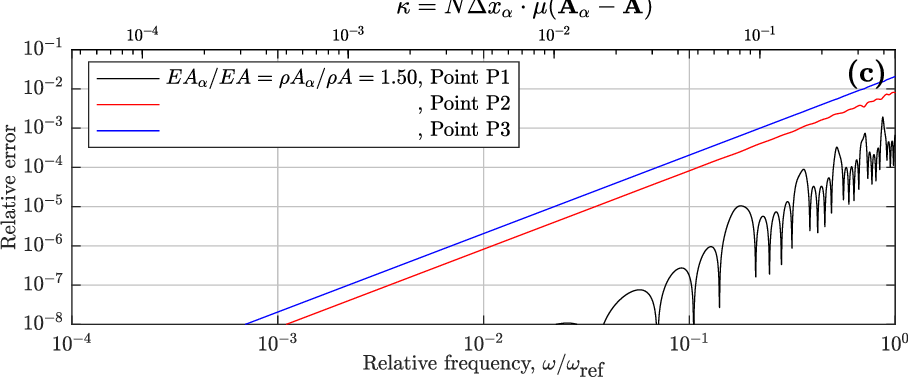} &
			\includegraphics[width=8cm]{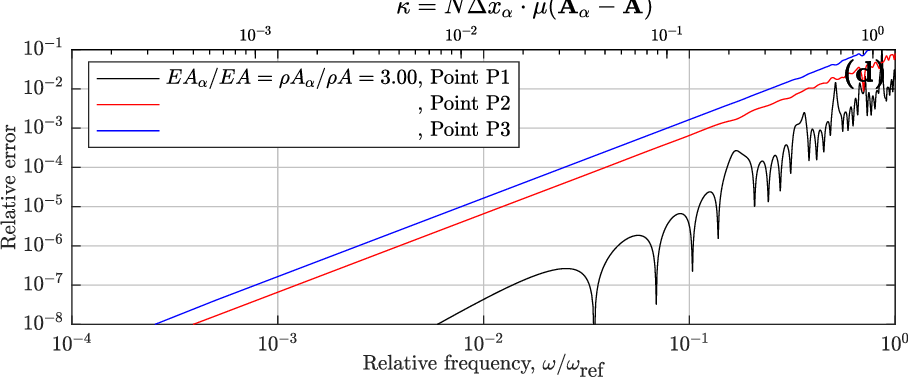} \\			
		\end{tabular}
}		
		\caption{Wavefield results of magnitude and error for an infinitely long rod containing $N=5$ periodically distributed scatterers, with a lattice spacing of $a = 0.4 \mathrm{m}$ under rightwards incident plane wave excitation. The contrast of properties considered is: $EA_\alpha / EA =\rho A_\alpha / \rho A = 1.50$ (left) and $EA_\alpha / EA =\rho A_\alpha / \rho A = 3.00$ (right). We consider the wavefield at the positions $P_1$, $P_2$ and $P_3$,  as defined in the schematic in the top panel where $L = 4 \mathrm{m}$. Panels (a) and (b) show the magnitude of wavefield, (c) and (d) show relative error as function of frequency and of the parameter $\kappa$.}%
		\label{fig03}%

\end{figure}
 We use the notation $\mu(\cdot)$ instead of the classical {\em rho}-notation $\rho(\cdot)$ for the spectral radius, to avoid confusion with the density. The parameter $\kappa$ is dimensionless since the eigenvalues of matrices $\mathbf{A}$ and  $\mathbf{A}_\alpha $ have physical meaning as wavenumbers. The relationship between this parameter and the relative error of the wavefield for the three points P1, P2 and P3 has been plotted in Figs. \ref{fig02}(d) and \ref{fig02}(f). The relative error is defined as 
 \begin{equation}
 \text{relative error} = \frac{\left|u(x,\omega)-u_{\textnormal{approx}}(x,\omega)\right|}{\left|u(x,\omega)\right|}  \, .
 \label{eq090}
 \end{equation}

As expected, the relative error of the approximation decreases as $\kappa$ becomes smaller with the logarithmic scale revealing the relative error is of $\mathcal{O}(\kappa^2)$ for any of the considered points (P1, P2 and P3). Therefore, Eq. \eqref{eq013} accurately predicts the behaviour of waves arriving to the scatterer array (point P1), where the field is the sum of the incident and the reflected waves. As seen in Figs. \ref{fig02}(d) and \ref{fig02}(f), the error accumulates as the wave passes each approximated scatterer. This behaviour can be generalized to other waveguides and for other configurations. In general, modelling small inclusions as point-source terms produces a waveform very close to that of the exact model, and the error in doing so is of $\mathcal{O}(\kappa^2)$ which increases  with increasing: $\omega$, $\Delta x_{\alpha}$ and with the contrast of properties. When changing $N$, $\Delta x_\alpha$ and $(\mathbf{A}_\alpha - \mathbf{A})$ independently but keeping $\kappa$ constant (as when comparing Fig. \ref{fig02}(d) with Fig. \ref{fig02}(f))  the relative error is found to be of the same order of magnitude. \\

\begin{figure}[H]%
\hspace*{-0.45cm}
\scalebox{0.9}{
		\begin{tabular}{cc}
			\includegraphics[width=8cm]{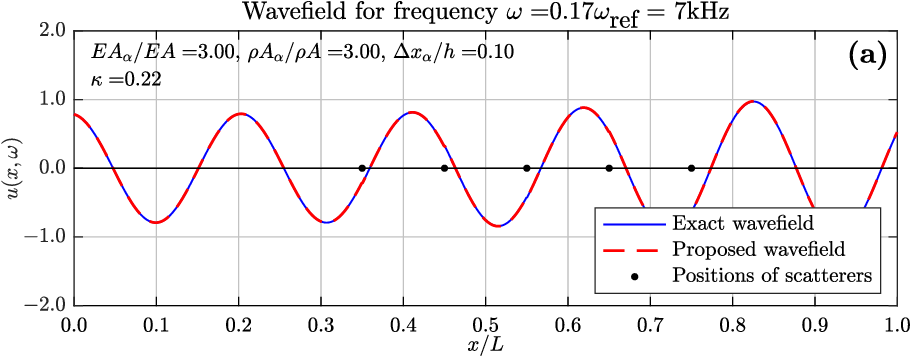}	&
			\includegraphics[width=8cm]{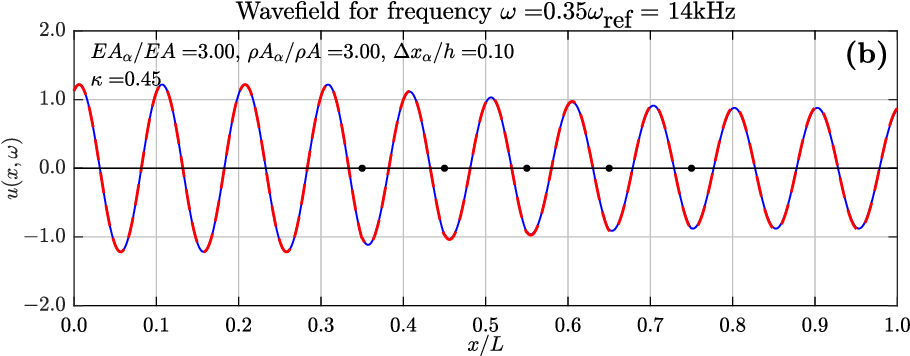} \\ \\		
			\includegraphics[width=8cm]{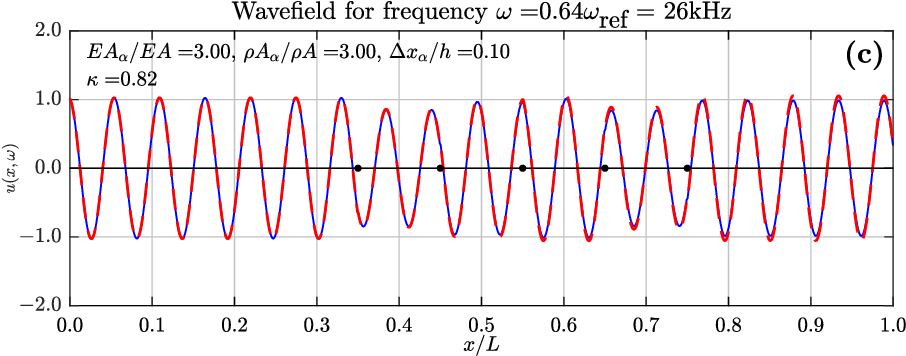}  & 	    
			\includegraphics[width=8cm]{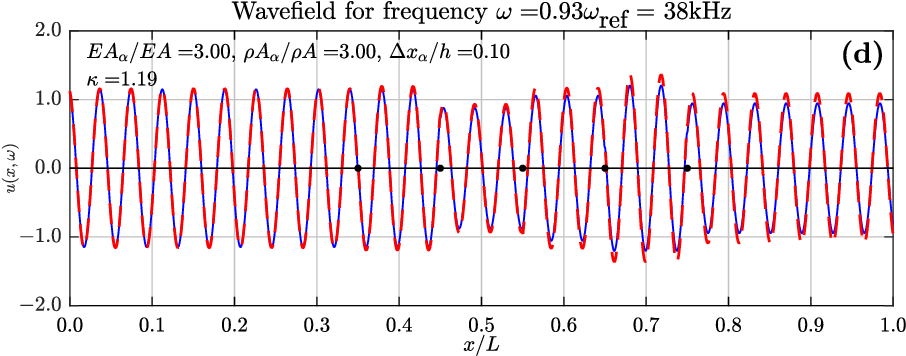} 			
		\end{tabular}
}		
		\caption{The displacement field of the configuration from Fig. \ref{fig03}, with parameters $EA_\alpha / EA =\rho A_\alpha / \rho A = 3$, $\Delta x_\alpha / h = 0.10$, subject to plane-wave excitation for four different frequencies. The parameter $\kappa$ changes for each waveform and is given in the top left corner of each graph.}%
		\label{fig04}%
\end{figure}

In Fig.~\ref{fig03} we consider the results for a configuration of inclusions arranged periodically in space. The number of scatterers is $N=5$ and $\Delta x_\alpha = 0.10h$, where $h = 5 \mathrm{cm}$ is the height of the rod. The rigidity and mass parameters of the inclusions are $EA_\alpha/EA = \rho A_\alpha / \rho A = 1.50 $ for Fig.~\ref{fig03}(a) and \ref{fig03}(c) and $EA_\alpha/EA = \rho A_\alpha / \rho A = 3.00 $ for Fig.~\ref{fig03}(b) and \ref{fig03}(d). The Bragg peaks of the periodic distribution arise at the following frequencies \cite{Hussein-2017}
\begin{equation}
	\omega_{\textnormal{Bragg}} = \frac{n \pi}{a} \sqrt{\frac{EA}{\rho A}} = 
	\{0.176 \, \omega_{\text{ref}}, \ 0.353 \, \omega_{\text{ref}}, 0.529 \, \omega_{\text{ref}}, \dots \} \, ,
	\label{eq087}
\end{equation}

\noindent where $a= 0.4$ m is the separation distance between scatterers. In Fig. \ref{fig03} we see a good agreement between the approximate and exact solutions for the wavefield over the entire frequency range  considered, even when considering a relatively high contrast between the host-medium and inclusions as in Fig.~\ref{fig03}(b) and \ref{fig03}(d). The plots of relative error depict the same pattern as observed in Fig. \ref{fig02}: again the reflected wave is more accurate (several orders of magnitude) than the transmitted wave for any frequency, and the error accumulates as the wave passes every approximated scatterer.  Again, the relative error is of $\mathcal{O}(\kappa^2 )$. \\

In Figs.~\ref{fig04}(a)-(d), we consider the same configuration in Fig. \ref{fig03} (b) \& (d) where now the stationary solution is plotted over a section of the waveguide for four different frequencies. In Figs.~\ref{fig04}(a)-(d), the approximate solution has remarkable accuracy along the whole length of the rod for all cases; however, observe that as we reduce the wavelength by increasing the frequency, a loss of precision is perceived downstream of the scatterers as seen by the comparisons of the wavefield at points P1, P2 and P3 in Fig. \ref{fig03}. Each one of the wavefields shows the value of $\kappa$ associated with the simulation. In general, after testing with a large number of simulations, values of $\kappa \le 1$ are associated with satisfactory results of the proposed method.

\section{Numerical example: multiple scattering from flexural waves in a Timoshenko beam} \label{NES2}

\begin{figure}[H]%
\hspace*{2.0cm}
\scalebox{0.95}{
		\begin{tabular}{cc}
			\includegraphics[width=10cm]{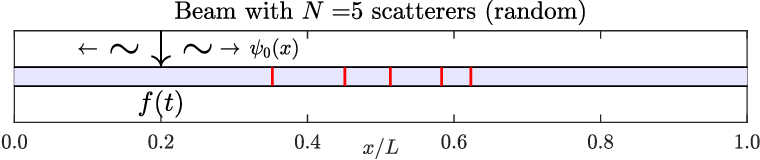} \\ \\
			\includegraphics[width=10cm]{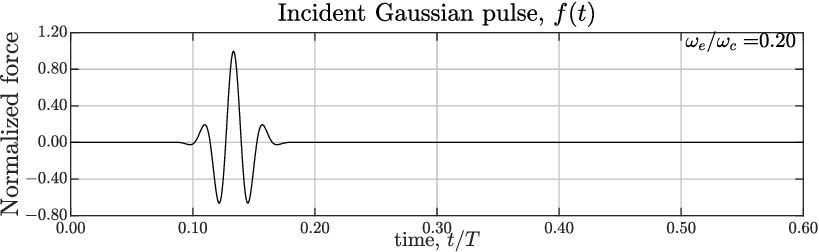} \\ \\
			\includegraphics[width=10cm]{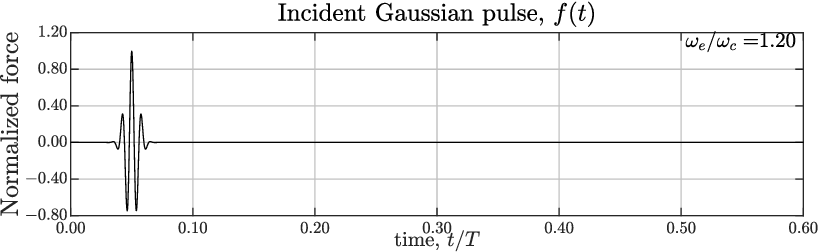}		
		\end{tabular}
}		
		\caption{(Top) An infinitely long beam containing $N=5$ randomly distributed scatterers under point source excitation $f(t)$, we consider the wavefield over a $L = 3.5$ m portion of the beam as shown. (Middle) Normalized time-domain Gaussian force-pulse for the low-frequency case 1, with a time-window of $T =15$ ms and frequency-pulse of $\omega_e = 0.20\omega_c$. (Bottom) Normalized time-domain Gaussian force-pulse for the high-frequency case 2, with a time-window of $T=8$ ms and frequency-pulse of $\omega_e = 1.20\omega_c$. Here  $\omega_c = \sqrt{GA_z/\rho I_y}= 13$ kHz.  }%
		\label{fig07}%
\end{figure}

In this section, the proposed approximation will be validated for the Timoshenko beam model which  considers both rotational and shear effects. Refer to Table \ref{tab03} for  $\mathbf{A}$ in the Timoshenko $2m=4$ case. We consider a aluminium beam with a 12$\times$12 cm$^2$ square cross section, resulting in a cut--off frequency of $\omega_c = \sqrt{GA_z/\rho I_y}= 13$ kHz for the homogeneous beam. The aim of this section is to study the time-domain wavefield propagation produced by a point-source excitation, $f(t)$, placed at the origin $x=0$. The excitation has a Gaussian-form pulse centred at $\omega = \omega_e$, where we consider two values for $\omega_{e}$ to validate the model: (1) the low-frequency case with $\omega_e = 0.2\omega_c$, and (2) the high-frequency case with $\omega_e = 1.2\omega_c$. For this example, a configuration of $N=5$ scatterers as shown in Fig. \ref{fig07} is used. For $\boldsymbol{\psi}_{0}$, we consider case $(ii)$ in section \ref{incFe}, and utilise the Fourier expansion in the applied frequency range \cite{Mace-1984, Mace-2005} to determine the field in the time-domain. \\

\begin{table}[H]	
	\begin{center}
		{\input{table_02.tex}}
		\caption{Mechanical properties of the scatterers with respect to those of the empty-beam. The two cases, low- and high-frequency refers to the main frequency of the excitation pulse $\omega_e = 0.20 \omega_c$ and $\omega_e = 1.20 \omega_c$ respectively.}
		\label{tab02}
	\end{center}	
\end{table}

The mechanical properties of the inclusions with respect to the empty--beam are listed in Table \ref{tab02} for the two cases considered. The corresponding value of the scattering parameter $\kappa$ has also been added to Table \ref{tab02}. The contrast of properties, number and width and inclusions are kept constant from one case to the other. Since case 2 is of a higher frequency than case 1, the scattering parameter $\kappa$ is one order of magnitude higher. Again, the approximate model from Eq. \eqref{eq013} will be tested against the exact solution from the TMM.  \\

\begin{figure}[H]%
\hspace*{1.75cm}
\scalebox{0.95}{
		\begin{tabular}{cc}
			\textbf{Case 1: low-frequency} & 
			\textbf{Case 2: high-frequency} \\ \\
			\includegraphics[width=5cm]{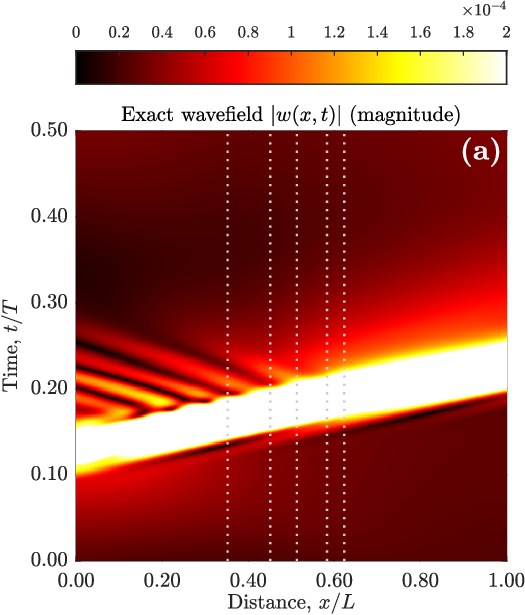} 	&
			\includegraphics[width=5cm]{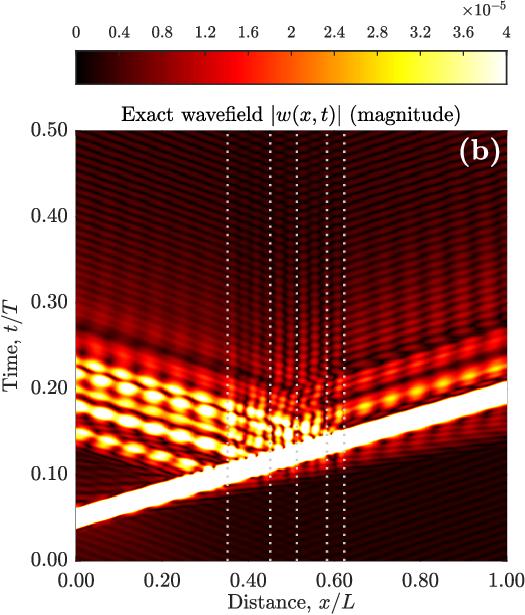} 	\\			\\
			\includegraphics[width=5cm]{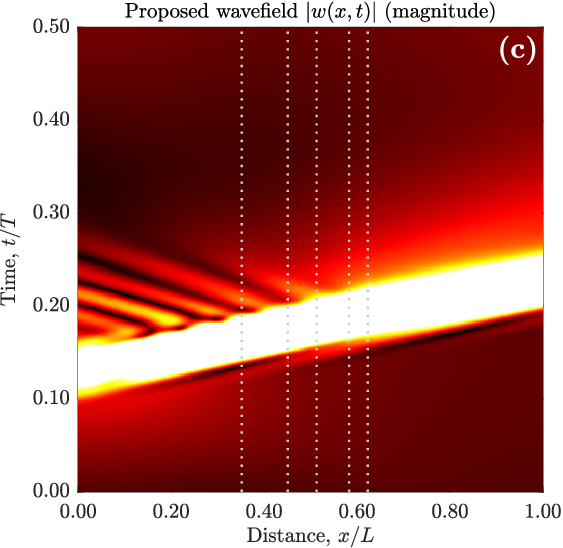}	&
			\includegraphics[width=5cm]{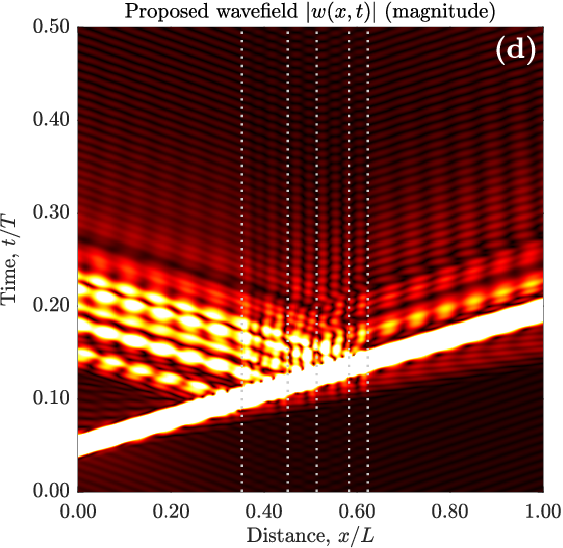}	\\				\\
			\includegraphics[width=5cm]{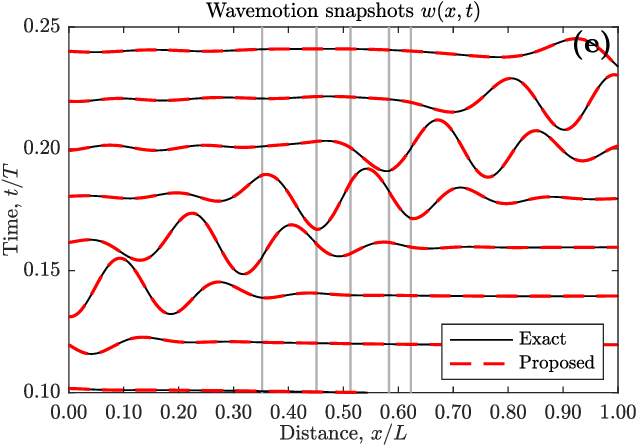}	&
			\includegraphics[width=5cm]{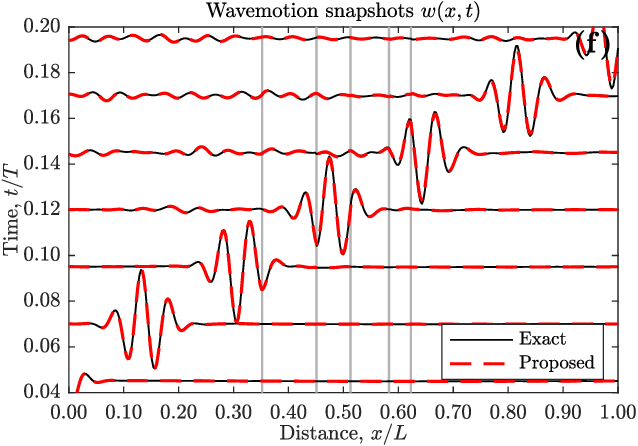}	\\	\\
		\end{tabular}
}
		\caption{Comparison between the exact and approximate solution for the time-domain simulations of the multiple scattering problem outlined in Fig. \ref{fig07}. Left plots show the space-time propagation results of case 1 (low-frequency excitation frequency  $\omega_e = 0.20 \omega_c$): (a) Exact wavefield magnitude, (c) Approximate wavefield magnitude, (e) Some simulations snapshots. 
		Right plots show the space-time propagation results of case 2 (high-frequency, excitation frequency  $\omega_e = 1.20 \omega_c$): (b) Exact wavefield magnitude, (d) Approximate wavefield magnitude, (f) Some simulations snapshots.}%
		\label{fig05}%
\end{figure}

In Fig. \ref{fig05} the results of the time-domain simulations are plotted. Case 1 has been plotted in the graphs of the left column: Figs. \ref{fig05}(a) and \ref{fig05}(c) depict the wavefield in magnitude for case 1 (low-frequency) for the exact and approximate solutions respectively. Figs. \ref{fig05}(b) and \ref{fig05}(d) show the same variables and methods but for case 2 (high-frequency).  \\

As in section \ref{NES1}, on comparing the exact and approximate simulations in Fig. \ref{fig05}, we see that the accuracy of the approximate solution drops off as the wavelength decreases. Again, the error accumulates as the wave passes every approximated scatterer. In the time-domain simulations it is seen that in the low frequency case, Figs. \ref{fig05}(a) and \ref{fig05}(c), the fit is almost perfect and no discrepancies are observed. For this case, the scattering parameter $\kappa = 0.176$ and is associated with very good results obtained by the approximate model. As the excitation frequency increases, the value of $\kappa$ increases, and in case 2 $\kappa = 1.06$. For smaller wavelengths, the small scatterer approximation by point--force terms is not as good, and this is observed when comparing Figs \ref{fig05}(b) and \ref{fig05}(d) or considering Fig. \ref{fig05}(f) - where the disagreement between the exact and the approximate solution is more evident than in the lower frequency case. However, similarly to section  \ref{NES1}, it should be noted that the loss of accuracy of the approximate wavefield is only observed in amplitude and not in the phase of the wave. \\

Observe, in Fig \ref{fig05}(b) and Fig \ref{fig05}(d), that the approximate solution still produces reasonable results with the correct behaviour; for instance, the shear modes still travel at a higher velocity than the bending modes. Moreover, it is remarkable that the approximate solution produces a wavefield that roughly matches that of the exact solution when the  $\kappa$ parameter is high. For smaller values of $\kappa$, it is evident that the approximate method accurately simulates structures under high frequency ranges so long as the other parameters (number of scatterers, width and contrast properties) are chosen appropriately. Moreover, after performing an extensive parameter sweep over many simulations, we can say that our proposed approximate method is accurate when $\kappa \leq 1$ for higher order beam models. It is clear that the distribution of inclusions has some influence on the error of the approximation, as the errors are not the  same if the $N$ obstacles are randomly distributed or placed periodically. However, for all multiple scattering simulations considered, the same scattering parameter $\kappa \le 1$ is a good indicator that the approximation used will be accurate. A more detailed analysis of the errors introduced by approximating small inclusions with point source terms is left for future work.

\section{Conclusions}

We studied the propagation of elastic waves in one-dimensional waveguides containing multiple obstacles in the form of inclusions. Both the empty-waveguide and the inclusions were considered to be formed from homogeneous materials. When considering the inclusions to be embedded within the empty-waveguide, we solved the multiple scattering problem when a change in the mechanical properties takes place in-between an inclusion and its surrounding material. Here, either changes in the cross section and/or the material parameters were considered. The size of the inclusions were considered small in comparison to the wavelength, and in such cases, through the formalism of generalised functions we demonstrated how these inclusions can be approximated by point--force source terms applied to the empty-waveguide. Subsequently, we applied this methodology to construct the approximate solutions for the multiple scattering problem through the Green's function of the empty-waveguide, and this was the main aim of the paper.  \\

The theoretical procedure was developed for the general case in first order form, where the constitutive relations appropriate for any waveguide were expressed as a matrix. The general solution was determined through the Green's function in matrix form, and is appropriate to consider multiple scattering problems for one-dimensional waveguides, modelled with an arbitrary number of degrees of freedom and containing an arbitrary number of inclusions.  These expressions are simple to utilize for multiple scattering simulations within different types of waveguides, this was demonstrated with two numerical examples: a rod with longitudinal waves and a Timoshenko beam with bending and shear waves. We cross-verified our approximate solutions against the exact solution using the TMM, and proposed a  dimensionless scattering parameter $\kappa$ to discuss quantitatively how strongly scattering the inclusions were and to estimate the order of the relative errors within the approximate solution. The comparison of results covers the frequency domain and the time domain verifying that, in general, there is a good agreement between the approximate and the exact results when $\kappa$ is less than or equal to unity. We anticipate that our proposed approximate method, based upon Green's functions,  is versatile and simple enough to consider coupled problems containing many degrees of freedom, with implications for the design of structured media to control wave propagation.

\section*{Acknowledgements}

M.L. and L.M.G.-R. are grateful for the partial support under Grant No. PID2020-112759GB-I00 funded by 
MCIN/AEI/10.13039/501100011033, and also for the partial support under Grant No. PID2023-146237NB-I00 funded by MCIU/AEI. L.M.G.-R. acknowledge support from 
Grant No. CIAICO/2022/052 of the ``Programa para la promoci\'on de la investigaci\'on cient\'ifica, el desarrollo tecnol\'ogico y la innovaci\'on en la Comunitat Valenciana'' funded by Generalitat Valenciana. M.L is grateful for support under the ``Programa de Recualificaci\'on del Sistema Universitario Espa\~nol para 2021-2023'', funded by ``Instrumento Europeo de Recuperaci\'on (Next Generation EU) en el marco del Plan de Recuperaci\'on, Transformaci\'on y Resiliencia de Espa\~na'', a trav\'es del Ministerio de Universidades. R.W and R.V.C acknowledge  financial support from the EU H2020 FET-proactive project MetaVEH under grant agreement number 952039.


\bibliographystyle{unsrtnat} 
\bibliography{bibliography,Bibli}

\appendix

\section*{Appendix}
\section{Examples of one-dimensional elastic waveguides}
\label{ap_C}

Refer to Table \ref{tab03} for the matrices $\textbf{A}$ governing the constitutive relations for different types of one-dimensional elastic waveguides. These matrices can readily be inserted into the expressions in Eqs ~\eqref{eq013}, \eqref{eq043} \& \eqref{eq073} to solve multiple scattering problems for any waveguide of interest.
%

\begin{table}[h]	
\scalebox{0.9}{
		\scriptsize{\input{table_03.tex}}
}		
		\caption{State--vector, force vector and matrix of parameters for four different types of elastic waveguides, those with: purely longitudinal waves, purely torsional waves, purely flexural waves and coupled problems.}
		\label{tab03}

\end{table}

\end{document}

%% file: table_01.tex


\begin{tabular}{c >{\centering\arraybackslash} m{1.5cm} >{\centering\arraybackslash} m{1.5cm} >{\centering\arraybackslash} m{6.5cm}m{1.5cm}}
	Model 				   		&			$\mathbf{u}(x)$	&			$\mathbf{f}(x)$ & Matrix $\mathbf{A}$	 &  \hspace*{-0.25cm} Sketch 	\\ 
	\hline \hline
	\begin{tabular}{l} Longitudinal \\ (classical rod) \\ $2m = 2$  	\end{tabular}	 
								 &			 $\left\{\begin{array}{cc} u \\ N_x \end{array} \right\}$
								 &			 $\left\{\begin{array}{cc} 0 \\ -p_x \end{array}  \right\}$ 
								 &			 $\left[ \begin{array}{cc} 0  & 1/EA \\ - \rho A \omega^2 & 0 \end{array} \right]$  
								 &			\hspace*{-0.5cm} \includegraphics[width=2.2cm]{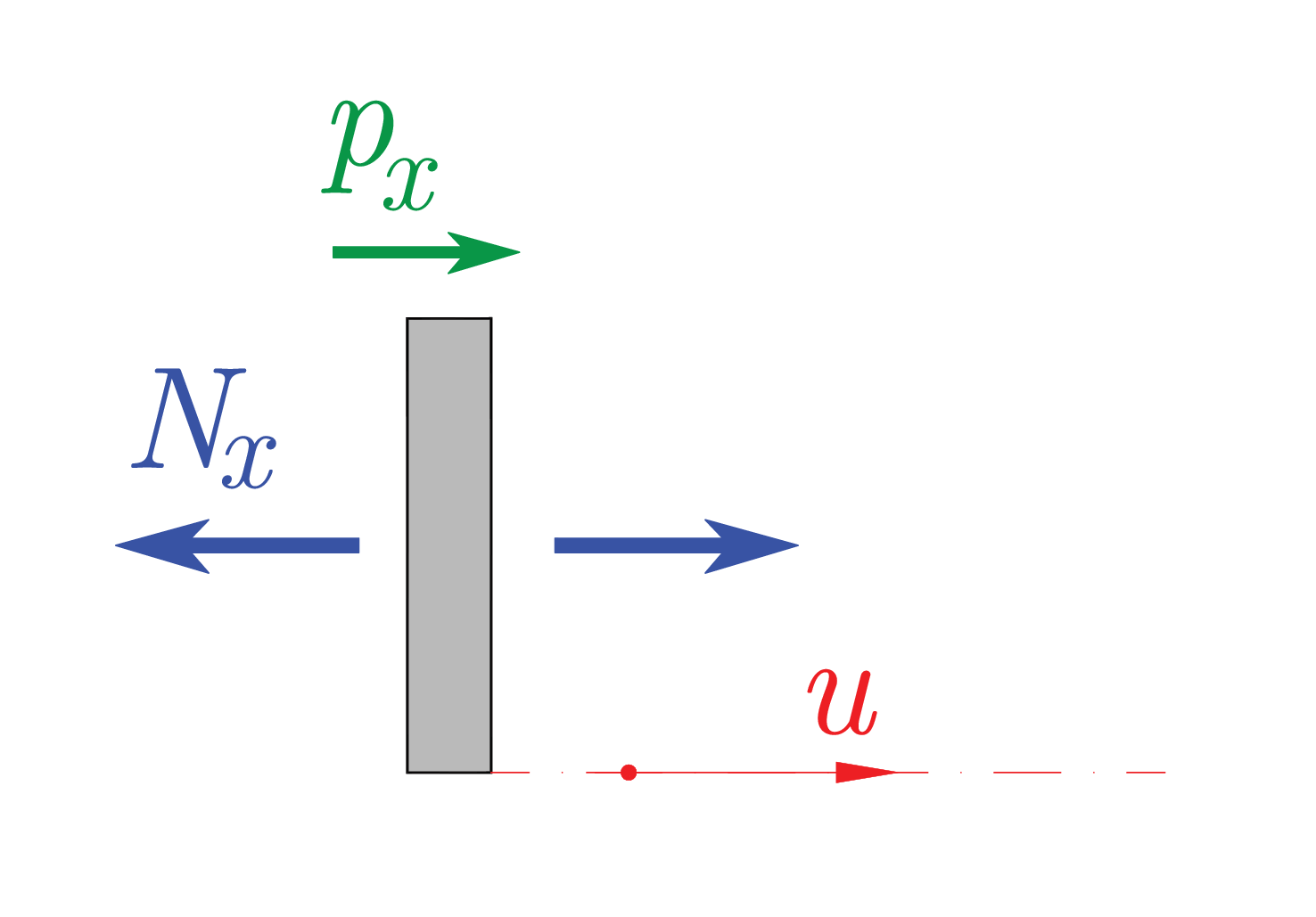} \\ \hline 
	\begin{tabular}{l} Flexural \\ (Euler-Bernoulli) \\  	$2m = 4$ \end{tabular} 	 
								&			 $\left\{\begin{array}{cc} w \\ \theta_y \\ V_z \\  M_y \end{array} \right\}$
								&			 $\left\{\begin{array}{cc} 0 \\ 0 \\ -p_z \\ - m_y \end{array}  \right\}$ 
								&			 $\left[ \begin{array}{cccc} 0  & 1 &  0 & 0 \\ 0 & 0 & 0 & 1/EI_y  \\ 
									- \rho A \omega^2 & 0 & 0 & 0 \\ 0 &  0  & -1 & 0  \end{array} \right]$ 
								& 		\hspace*{-0.5cm}	\includegraphics[width=2.2cm]{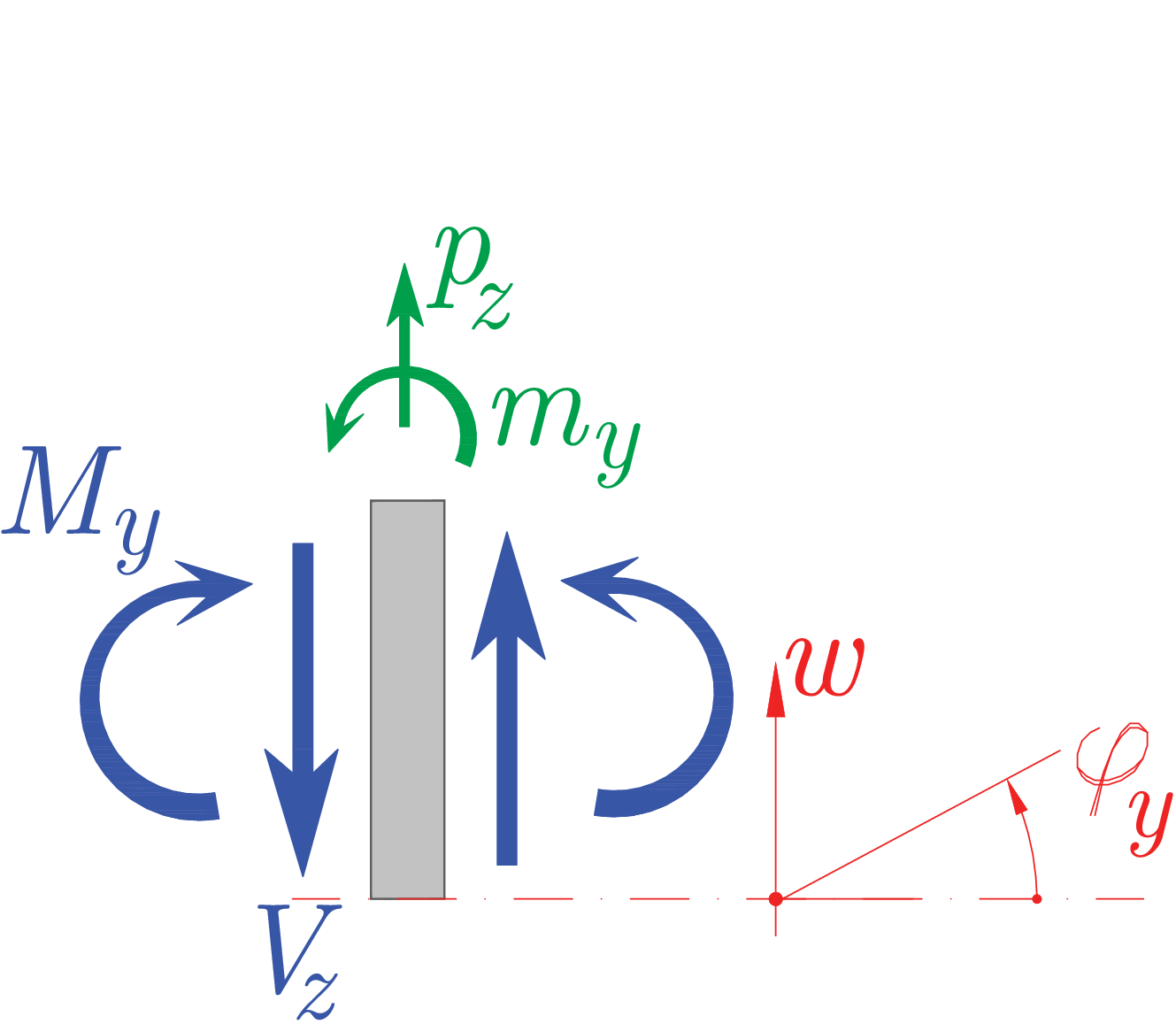}	\\	\\																			
\hline 																								
\end{tabular}

%% file: table_02.tex

\begin{tabular}{ccccc}
																			&    Case 1 (low-frequency)			&  Case 2 (high-frequency)			\\
	\hline																			
	Frequency, $\omega / \omega_c$		  &   										0.200       & 										1.200	\\
	$EI_\alpha/EI$  										&  										 0.512	  & 										0.512		\\
	$GA_\alpha / GA$									 &										 0.800		&										0.800				\\
	$\rho A_\alpha / \rho A$						&										 1.200	&										1.200		\\
	$\rho I_\alpha / \rho I$							&										0.768	&										0.768		\\	
	$\Delta x_\alpha / h$							   &										0.220       &									    0.220		\\
	$\kappa = N \, \Delta x_\alpha \cdot \mu \left(\mathbf{A}_\alpha - \mathbf{A}\right) $
																			&										\textbf{0.1766}		&										\textbf{1.059}  \\
	\hline
\end{tabular}

%% file: table_03.tex


\begin{tabular}{lccccr}
	Model 				   		&			$\mathbf{u}(x)$	&			$\mathbf{f}(x)$ & Matrix $\mathbf{A}$	 	\\ 
	\hline \\
	\begin{tabular}{l} Longitudinal \\ (classical rod) \\ $2m = 2$  	\end{tabular}	 
								 &			 $\left\{\begin{array}{cc} u \\ N_x \end{array} \right\}$
								 &			 $\left\{\begin{array}{cc} 0 \\ -q_x \end{array}  \right\}$ 
								 &			 $\left[ \begin{array}{cc} 0  & 1/EA \\ - \rho A \omega^2 & 0 \end{array} \right]$  \\ \\
	\begin{tabular}{l} Longitudinal \\ (Love's rod)  	\\ $2m=2$ \end{tabular}	 
								&			 $\left\{\begin{array}{cc} u \\ N_x \end{array} \right\}$
								&			 $\left\{\begin{array}{cc} 0 \\ -q_x \end{array}  \right\}$ 
								&			 $\left[ \begin{array}{cc} 0  & 1/(EA-\rho I_x \nu^2 \omega^2) \\ - \rho A \omega^2 & 0 \end{array} \right]$  \\ \\
	\begin{tabular}{l} Torsional  \\ (Saint-Vennat) \\ $2m = 2$  	\end{tabular}	 
								&			 $\left\{\begin{array}{cc} \theta_x  \\ T_x \end{array} \right\}$
								&			 $\left\{\begin{array}{cc} 0 \\ -m_x \end{array}  \right\}$ 
								&			 $\left[ \begin{array}{cc} 0  & 1/GJ \\ - \rho I_x \omega^2 & 0 \end{array} \right]$  \\ \\		
	\begin{tabular}{l} Torsional  \\ (Vlassov) \\ $2m = 4$ 	\end{tabular} 	 
								&			 $\left\{\begin{array}{cc} \theta_x \\ \varphi \\ T_x \\  M_w \end{array} \right\}$
								&			 $\left\{\begin{array}{cc} 0 \\ 0 \\ -m_x  \\ 0 \end{array}  \right\}$ 								
								&			 $\left[ \begin{array}{cccc} 0  & 1 &  0 & 0 \\ 0 & 0 & 0 & 1/EI_w  \\ 
									- \rho I_x \omega^2 & 0 & 0 & 0 \\ 0 &  \rho I_w \omega^2 - GJ  & -1 & 0  \end{array} \right]$ 		\\ \\														
	\begin{tabular}{l} Flexural \\ (Euler-Bernoulli) \\  	$2m = 4$ \end{tabular} 	 
								&			 $\left\{\begin{array}{cc} w \\ \theta_y \\ V_z \\  M_y \end{array} \right\}$
								&			 $\left\{\begin{array}{cc} 0 \\ 0 \\ -q_z \\ - m_y \end{array}  \right\}$ 
								&			 $\left[ \begin{array}{cccc} 0  & 1 &  0 & 0 \\ 0 & 0 & 0 & 1/EI_y  \\ 
																		- \rho A \omega^2 & 0 & 0 & 0 \\ 0 &  0  & -1 & 0  \end{array} \right]$ 	\\	\\			
	\begin{tabular}{l} Flexural \\ (Timoshenko) \\  	$2m = 4$ \end{tabular} 	 
								&			 $\left\{\begin{array}{cc} w \\ \theta_y \\ V_z \\  M_y \end{array} \right\}$
								&			 $\left\{\begin{array}{cc} 0 \\ 0 \\ -q_z \\ - m_y \end{array}  \right\}$ 
								&			 $\left[ \begin{array}{cccc} 0  & 1 &  1/GA_z & 0 \\ 0 & 0 & 0 & 1/EI_y  \\ 
									- \rho A \omega^2 & 0 & 0 & 0 \\ 0 &  -\rho I_y \omega^2  & -1 & 0  \end{array} \right]$ 	\\	\\																							
	\begin{tabular}{l} Flexural-torsional \\ $y$--symmetry \\ (Timoshenko, \\ Saint-Venant) \\ $2m = 6$	\end{tabular} 	 
								&			 $\left\{\begin{array}{cc} w \\ \theta_y \\ \theta_x \\ V_z \\ M_y \\  T_x \end{array} \right\}$
								&			 $\left\{\begin{array}{cc} 0 \\ 0 \\ 0 \\ -q_z \\ - m_y \\ - m_x  \end{array}  \right\}$ 
								&			 $\left[ \begin{array}{cccccc} 0  & 1 &  0 & 1/GA_z & 0 & 0 \\ 0 & 0 & 0 &  0 &  1/EI_y & 0  \\ 
																			0 & 0 & 0 & 0 & 0 & 1 / GJ \\
																		- \rho A \omega^2 & 0 & - \rho A y_G \omega^2 & 0 & 0 & 0 \\ 
																			0 &  - \rho I_y \omega^2 & 0   & -1 & 0 & 0 \\
																			- \rho A y_G \omega^2 & 0 &   - \rho I_x \omega^2 & 0 & 0 & 0 \end{array} \right]$ 					\\ \\ \hline 																								
\end{tabular}